\documentclass[]{aastex62}
\shorttitle{The Cygnus Loop with LAMOST}
\shortauthors{Seok et al.}

\defcitealias{fesen82}{F82}
\defcitealias{OF06}{OF06}

\def\vsh{$v_{\rm s}$}
\def\kms{${\rm km~s}^{-1}$}
 
\begin{document}

\title{Unbiased Spectroscopic Study of the Cygnus Loop with LAMOST. I.\\Optical Properties of Emission Lines and the Global Spectrum}

\correspondingauthor{Ji Yeon Seok}
\email{jyseok@kasi.re.kr}

\author{Ji Yeon Seok}
\altaffiliation{LAMOST Fellow}
\affiliation{Key Laboratory of Optical Astronomy, National Astronomical Observatories, Chinese Academy of Sciences,
Beijing 100010, China}
\affiliation{Korea Astronomy and Space Science Institute, 
Daejeon, 305-348, Korea}

\author{Bon-Chul Koo}
\affiliation{Seoul National University, 
Seoul 151-742, Korea}

\author{Gang Zhao}
\affiliation{Key Laboratory of Optical Astronomy, National Astronomical Observatories, Chinese Academy of Sciences,
Beijing 100010, China}
\affiliation{School of Astronomy and Space Science, University of Chinese Academy of Sciences, 
Beijing 100049, People’s Republic of China}

\author{John C. Raymond}
\affiliation{Harvard-Smithsonian Center for Astrophysics, 
60 Garden St., Cambridge, MA 02138, USA}

\begin{abstract}
We present an unbiased spectroscopic study of the Galactic supernova remnant (SNR) Cygnus Loop using the LAMOST DR5. LAMOST features both a large field-of-view and a large aperture, which allow us to simultaneously obtain 4000 spectra at $\sim$3700--9000~\AA~with $R\approx1800$. The Cygnus Loop is a prototype of middle-aged SNRs, which has advantages of being bright, large in angular size, and relatively unobscured by dust. Along the line of sight to the Cygnus Loop, 2747 LAMOST DR5 spectra are found in total, which are spatially distributed over the entire remnant. This spectral sample is free of the selection bias of most previous studies, which often focus on bright filaments or regions bright in [\ion{O}{3}]. Visual inspection verifies that 368 spectra (13$\%$ of the total) show clear spectral features to confirm their association with the remnant. In addition, 176 spectra with line emission show ambiguity of their origin but have a possible association to the SNR. In particular, the 154 spectra dominated by the SNR emission are further analyzed by identifying emission lines and measuring their intensities. We examine distributions of physical properties such as electron density and temperature, which vary significantly inside the remnant, using theoretical models. By combining a large number of the LAMOST spectra, a global spectrum representing the Cygnus Loop is constructed, which presents characteristics of radiative shocks. Finally, we discuss the effect of the unbiased spectral sample on the global spectrum and its implication to understand a spatially unresolved SNR in a distant galaxy.

\end{abstract}



\section{Introduction} \label{sec:intro}

One of the commonalities that supernova remnants (SNRs) have shown is that their structures and physical properties are nonuniform and inhomogeneous \citep[e.g.,][]{williams99, lopez11, seok13}. Diversity of SNR morphologies revealed by multi-waveband observations \citep[e.g.,][]{levenson95,rho98,rho01,reach02,hines04,koo16,yamane18} implies that their physical properties including temperature and density could strongly vary from one part to another even within a single SNR unlike the spherical symmetry that theoretical models for SNR evolution often assume \citep[e.g.,][]{chevalier74, martinez11}. The spatial variation of the physical properties inside an individual SNR is closely related to supernova (SN) explosion mechanisms \citep[e.g.,][]{hwang04, lopez11,peters13} as well as its surrounding environment \citep[e.g.,][]{chu97,bilikova07,lee12}. In particular for evolved SNRs, the latter plays a significant role for characterizing the nature of each SNR. 

Depending on the environment, various shock waves can be driven by SN explosions. When the ambient medium has a low density ($\leq$1 cm$^{-3}$), such shocks are usually (non-radiative) collisionless \citep[e.g.,][]{raymond91, draine93}. If a collisionless shock encounters (partially) neutral preshock gas, the optical emission from the shock is dominated by hydrogen emission lines, and it is referred to as ``Balmer-dominated'' \citep{chevalier78}. When a shock has accumulated a sufficient column density ($N_{\rm H}$), the energy loss via radiative cooling becomes significant. Then, the shock wave is referred to as a radiative shock. If a shock has not yet propagated enough to become fully radiative, the shock wave is incomplete (or truncated), with spectral features that differ from the emission spectrum of radiative shocks \citep{raymond88}. Such different types of shocks have been observed in SNRs and even inside a single SNR sometimes \citep[e.g., see][and references therein]{mckee80,raymond91,draine93,ghavamian13}.

The Cygnus Loop (G74.0--8.5) is a prototypical middle-aged SNR \citep[1.7--2.5$\times10^4$ yr,][]{miyata94,levenson98,fesen18}, which is among the brightest in optical and best-studied Galactic SNRs over the whole electromagnetic spectrum (e.g., gamma-ray: \citealp{katagiri11}, X-ray: \citealp{graham95, levenson97, levenson99, uchida09}, ultraviolet: \citealp{danforth00, seon06, kim14}, optical: \citealp{miller74, levenson98, blair05}, infrared: \citealp{braun86, arendt92, sankrit10}, radio: \citealp{leahy97, leahy98, leahy02, uyaniker02, uyaniker04}). It is large in angular size, covering nearly 3$\degr\times4\degr$ of the sky. The distance to the Cygnus Loop has been uncertain; Previous estimates range between $\sim400$ pc and 1 kpc, and the most recent estimate is $735\pm25$ pc based on Gaia parallaxes of three stars toward the remnant \citep[and references therein]{fesen18}. Adopting 735 pc, the physical size of the Cygnus Loop corresponds to $\sim38\times51$ pc. Despite local variations in the morphology observed in different wavelengths, the overall remnant has a complete shell with the breakout to the south. 

Taking advantage of its great extent, proximity, and relatively low interstellar extinction \citep{parker67,fesen82}, detailed structures associated with diverse types of shock waves inside the Cygnus Loop have been detected and examined \citep[e.g.,][]{miller74, raymond80, raymond88, fesen82, levenson98, blair05, sankrit14}. In particular, a few selected locations including the prominent emission regions such as the Eastern and Western Veil Nebulae (NGC 6992 and NGC 6960, respectively) and the southernmost part of NGC 6992, the so called ``XA'' region \citep{hester86}, have been extensively investigated by using imaging as well as spectroscopy \citep[e.g.,][]{raymond88, hester94, levenson96, danforth01, blair05, medina14}. Bright optical emission in these regions arises from (complete or incomplete) recombination zones behind shock waves with a velocity of \vsh$\la100$ \kms~\citep[e.g.,][]{raymond88} whereas faint Balmer-dominated filaments often found outside the bright emission regions are produced by a fast, nonraditive shock with a velocity of \vsh$\ga150$ \kms~\citep[e.g.,][]{blair05}.

To understand the evolution of the Loop and its large-scale influence on the ambient medium comprehensively, it is essential to examine physical (and chemical) properties of the entire remnant such as shock velocities, preshock densities, and abundances and take their spatial variations into account. In general, spectral and spatial information from a remnant can be obtained by performing spectral mapping or integral-field spectroscopy. Such an approach is, however, limited to small objects in angular size or one portion of a large object. For a large object like the Cygnus Loop, it is practically unfeasible to obtain spectra of the entire region in the same manner. Consequently, previous studies with optical spectroscopy often either have focused on specific regions \citep[e.g.,][]{raymond88, danforth01, patnaude02} or have collected spectra from a few positions \citep[e.g.,][]{miller74, fesen82}. On the other hand, multi-object spectroscopy with a large field-of-view can provide an alternative way to evaluate global properties efficiently. Recently, \citet{medina14} have used a multi-object echelle spectrograph, Hectochelle mounted at the MMT 6.5 m telescope to examine collisionless shocks in the northeast limb of the Loop. High-resolution spectra covering H$\alpha$ and [\ion{N}{2}] $\lambda\lambda$6548, 6583 ($\sim6460$--6670 \AA) were obtained from 240 locations inside 1$\degr$ region, which allowed them to constrain properties of both the pre-shock and post-shock gas around Balmer-dominated filaments. 

In this paper, we present the first results of the extensive, multi-object spectroscopic observations carried out toward the entire region ($4\degr\times4\degr$) of the Cygnus Loop using the Large Sky Area Multi-Object Fiber Spectroscopic Telescope (LAMOST) Data Release 5 (DR 5). Section \ref{sec:data} describes a brief overview of the LAMOST data, selection of spectra associated with the remnant, and line identification. In Section \ref{sec:res}, we examine line ratios and their mutual correlations and derive physical properties. Then, we construct a global spectrum of the Cygnus Loop and discuss its global characteristics and implications for extragalactic SNRs in Section \ref{sec:global}. Finally, we summarize the main results in Section \ref{sec:sum}. Detailed analysis of kinematics, spatial variation, and shock modelling will be discussed in forthcoming papers.


\section{Data} \label{sec:data}

\subsection{LAMOST data}  \label{subsec:lamost}

We have examined the Cygnus Loop using spectra from LAMOST DR 5 released on December 2017. LAMOST (also known as Guo Shou Jing Telescope) features both a wide field of view ($\sim20$ deg$^2$) as well as a large aperture ($\sim4$ m in diameter), and 16 spectrographs equipped with 32 4K$\times$4K CCDs allow us to obtain 4000 spectra simultaneously \citep{cui12}. Blue (3700--5900 \AA) and red (5700--9000 \AA) spectra are recorded separately with two CCDs. The spectral coverage is 3700--9000 \AA, and a spectral resolution of $R\approx1800$ (corresponding a velocity resolution of $\sim167$ km s$^{-1}$) is achieved by placing slit masks of 2/3 width of the fibers \citep[i.e., $2\farcs2$ in diameter;][]{zhao12,luo15}. LAMOST raw data are reduced with the LAMOST 2D pipeline \citep{luo15}, which are similar to those of the Sloan Digital Sky Survey \citep{stoughton02}. The LAMOST 2D pipeline include basic pre-processing such as dark and bias subtraction, flat-fielding, and sky subtraction. The final output of the LAMOST data (combining blue and red channels) are one-dimensional relative flux-calibrated spectra. The data presented in this work are reduced using version 2.9.7 of the pipeline and can be directly downloaded from the LAMOST DR5 archive.\footnote{http://dr5.lamost.org/}. For spectra with a high signal to noise ratio (S/N) (i.e., S/N$\ge30$ at 4350 \AA), a precision of about 10 per cent between 4100 and 9000 \AA~is generally expected according to a comparison of the spectra of common objects obtained on different nights \citep{xiang15}. In this paper, we adopt 10$\%$ calibration error, and the final uncertainties are the quadratic sum of the calibration errors and the flux uncertainties mainly arising from the baseline fluctuation in gaussian fitting (see Section \ref{subsec:line}).

\begin{figure}[hbt!]
    \epsscale{1.2}
    \plotone{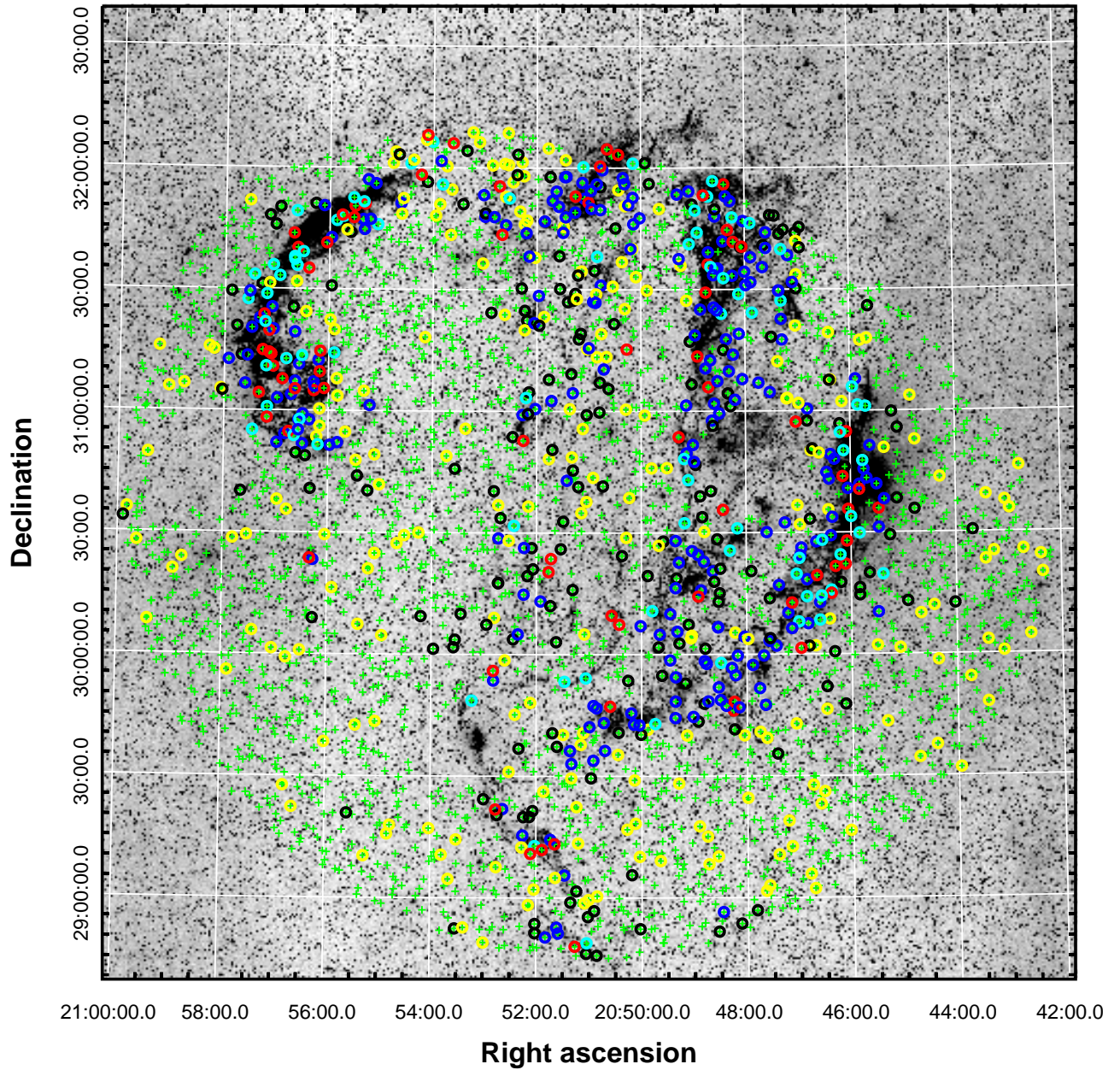}
    \caption{The Cygnus Loop reproduced from the red image of the DSS2 showing the locations of the 2747 LAMOST fibers (green crosses). Those selected for visual inspection (778 spectra, marked with circles) are classified into four groups: I. SNR-dominated spectra (red), II. SNR+stellar spectra (IIa: cyan, IIb: blue), III. stellar spectra with tentative SNR emission or ambiguous association with the SNR (black), and IV. no association with the SNR (yellow). 75, 79, 214, 176, and 234 spectra are included in Group I, IIa, IIb, III, and IV, respectively. }
    \label{fig:dss2}
\end{figure}

The field of the Cygnus Loop is included in one of the LAMOST regular surveys, the LAMOST Experiment for Galactic Understanding and Exploration \citep[LEGUE;][]{deng12}. 2747 LAMOST DR5 spectra are found in the direction of the Cygnus Loop centered at ($\alpha_{\rm J2000}$, $\delta_{\rm J2000}$)=(20$\rm^h$ 51$\rm^m$, +30$\degr$ 40$\arcmin$), which are evenly distributed over the entire SNR as shown in Figure \ref{fig:dss2}. The spectra were obtained in two separate dates, of which details are summarized in Table \ref{tab:obs}.

\begin{deluxetable}{ccCccc}[hbt!]
\tablecaption{Observation Summary \label{tab:obs}}
\tablecolumns{6}
\tablewidth{0pt}
\tablehead{
\colhead{Obs. date\tablenotemark{a}} &
\colhead{Plan ID} & \colhead{Seeing\tablenotemark{b}} & 
\colhead{Exposure Time} & \colhead{Number of spectra} \\
\colhead{(YYYY-mm-dd)} & \colhead{} &
\colhead{(arcsec)} & \colhead{(sec)} & \colhead{(count)}
}
\startdata
2016-09-30 & HD205307N293856B01 & 3\farcs2 & 4500 & 1543 \\
2016-11-02 & HD205307N293856V01 & 2\farcs6 & 1800 & 1204 \\
\enddata
\tablenotetext{a}{The observation median UTC}
\tablenotetext{b}{FWHM of point-spread function measured during exposure representing the weather condition at a given date}
\end{deluxetable}

Because all observations are a part of the LEGUE survey primarily targeting stars, they do not particularly aim to observe the SNR itself. For this reason, many of the spectra can possibly contain stellar emission (see below). To discriminate between spectra of the Cygnus Loop and those of other objects, we first screen all spectra automatically based on the presence of emission lines (i.e., [\ion{O}{3}] $\lambda5007$, H$\alpha$, and [\ion{S}{2}] $\lambda\lambda$6717, 6731), considering that most stellar spectra do not exhibit emission lines except peculiar stellar types such as Be star, Herbig Ae/Be star, and Wolf-Rayet star. By comparing the mean intensity at wavelength ranges of the emission lines and its adjacent continuum level, 778 spectra show the mean intensity is greater than the continuum level for one emission line or more. Then, we perform visual inspection to classify them into four groups; I) SNR-dominated spectra, II) SNR+stellar spectra, III) stellar spectra with tentative SNR emission or ambiguous association with the SNR, and IV) spectra not associated with the SNR. For Group II, we divide them into two subgroups, IIa and IIb: IIa spectra exhibit as rich emission lines as Group I shows, whereas only a few lines (H$\alpha$, [\ion{N}{2}], and [\ion{S}{2}] lines in most cases) are clearly detected in IIb spectra. Therefore, Group I and IIa spectra are mainly used for further analysis, yet Group IIb spectra are included when only [\ion{S}{2}] doublets are analyzed such as deriving electron densities ($n_{\rm e}$, see Section \ref{subsec:Te_ne}). Finally, the numbers of spectra in Groups I, IIa, IIb, III, and IV are 75, 79, 214, 176, and 234, respectively, which are marked with different colors in Figure \ref{fig:dss2}. Details of the spectrum classification are summarized in Table \ref{tab:specgr}.

\begin{figure}
    \epsscale{1.2}
    \plotone{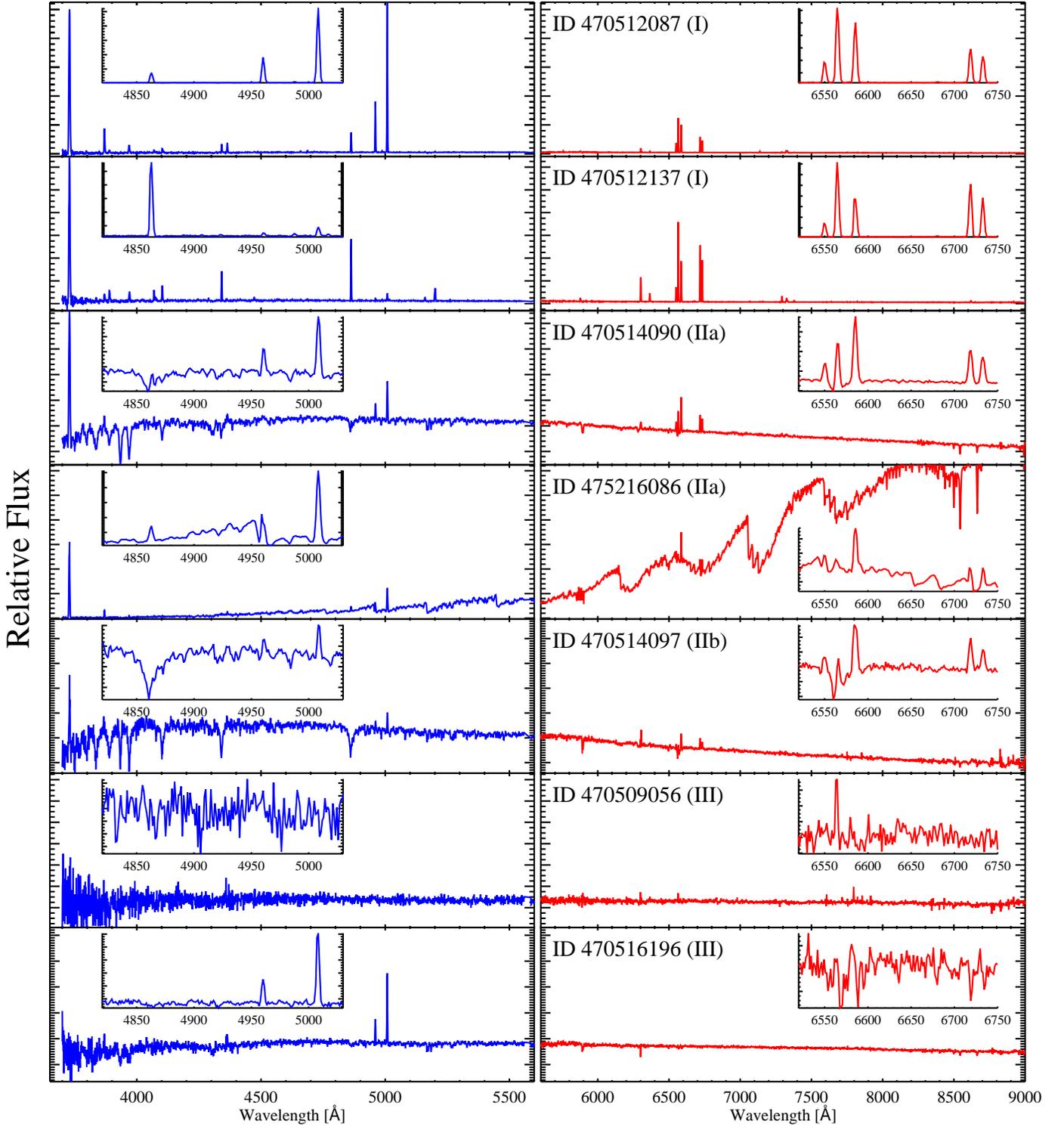}
    \caption{Exemplary LAMOST spectra of Group I, IIa, IIb, and III. Blue and red spectra are shown in left and right panels, respectively. Obs ID with its spectral group in parenthesis (see Table \ref{tab:specgr}) is marked in the right panel. In each panel, its zoomed-in spectrum near H$\beta$ or H$\alpha$ is displayed. The first and second spectra from top are representative of SNR-dominated Group I spectra with high and low [\ion{O}{3}]/H$\beta$ ratios, respectively. The two IIa spectra clearly show both SNR-related emission and stellar features. Superposed stars are likely to be F and M type (third and fourth rows, respectively). A spectrum in Group IIb (fifth row) exhibits some emission lines with strong absorption features. Two Group III spectra show limited sets of emission lines, and their origins are inconclusive. While only H$\alpha$ line appears in obs. ID 470509056 spectrum (sixth row), [\ion{O}{3}] lines with weak [\ion{O}{2}] $\lambda$3727 are present in obs. ID 470516196 spectrum (bottom row).} 
    \label{fig:EXspe}
\end{figure}

Group I contains 75 spectra dominated by SNR emission; Strong emission lines such as H$\alpha$, [\ion{S}{2}], [\ion{N}{2}], and [\ion{O}{3}] appear clearly, and stellar features such as absorption or continuum, if present, are negligible when SNR emission lines are extracted. Their associations with the Cygnus Loop are also confirmed by their spatial correspondences to the optical emission from the SNR seen in Figure \ref{fig:dss2}. Two exemplary spectra in Group I (obs. ID 470512087 and 470512137) are presented in Figure \ref{fig:EXspe}. The spectrum of obs. ID 470512087 is dominated by the strong [\ion{O}{3}] $\lambda\lambda$4959, 5007 lines whereas that of obs. ID 470512137 features the Balmer-series lines (i.e., H$\alpha$, H$\beta$, H$\gamma$, etc.).

Those showing both SNR emission and stellar features are classified into Group II. This can occur when diffuse emission from the Cygnus Loop and a background or foreground star are included within a single fiber. More than one third of the first-screened spectra (IIa+IIb: 293) belong to this category, which is a natural consequence considering the fact that the LAMOST survey primarily intends to target stellar objects in this field. Group II spectra show clear emission lines from the Cygnus Loop as well as non-negligible stellar features such as a series of hydrogen absorption features, \ion{Na}{1} and \ion{Ca}{2} absorption, and blue (or red) stellar continuum. In Figure \ref{fig:EXspe}, two IIa spectra are shown: obs. ID 470514090 and 475216086. The former spectrum shows the emission lines from the SNR on top of a F-type stellar spectrum featured by strong H and K of \ion{Ca}{2} whereas the latter shows the SNR emission as well as a M-type stellar spectrum characterized by a set of TiO bands. For this group, careful line identification is required, especially for those lines affected by strong absorption (see section \ref{subsec:line}). Locations of Group II spectra (IIa and IIb marked by cyan and blue circles, respectively in Figure \ref{fig:dss2}) are spatially in a good agreement with the SNR emission shown in the DSS2 image. 

Group III spectra exhibit emission lines of which origin is unclear. When several emission lines not usually seen in stellar spectra are marginally detected, these spectra are classified into this group. In some cases, Group III spectra show a few strong emission lines such as H$\alpha$, [\ion{S}{2}] $\lambda\lambda$6717,6731, or [\ion{O}{3}] $\lambda$5007. However, they do not have high [\ion{S}{2}]/H$\alpha$ ratios which is often used as a diagnostic of SNR origin \citep[e.g., see][]{mckee80,fesen85, long85}, or this ratio cannot be measured properly because H$\alpha$ or [\ion{S}{2}] lines (or both) are contaminated or do not appear. Moreover, their spatial correspondences to the SNR emission shown in the DSS2 image are not often discernible. Ambiguity of their association is partially due to lack of narrow band images (e.g., H$\alpha$ image). The presence of Balmer-dominated filaments from non-radiative shocks in the Cygnus Loop is well-known \citep[e.g.,][]{raymond83, fesen85, long92, hester94, sankrit00, ghavamian01, blair05, medina14, katsuda16}, but these filaments are usually fainter than emission from radiative shocks. Since faint Balmer-dominated filaments may not be distinct in the DSS2 red image, it is currently inconclusive whether those showing Balmer lines only in Group III originate from the Cygnus Loop. For example, a spectrum of obs. ID 470509056 presents a weak H$\alpha$ line only (see Figure \ref{fig:EXspe}). This spectrum might be related to one of Balmer-dominated filaments, but its location ($\alpha_{\rm J2000}$, $\delta_{\rm J2000}$)=(20:57:39.9, +30:40:05.7), relatively far away from the bright filaments, requires a further verification. Likewise, no prominent lines but [\ion{O}{3}] $\lambda\lambda$4959,5007 (and weak [\ion{O}{2}] $\lambda$3727) are found in several Group III spectra, which are most likely associated with the SNR, too. For example, the spectrum of obs. ID 470516196 shows (see the zoomed-in spectrum near H$\alpha$ in Figure \ref{fig:EXspe}), the presence of weak absorption at the wavelengths of [\ion{S}{2}] (and [\ion{N}{2}]), which implies that several emission lines from the SNR are oversubtracted during removal of night sky lines. There are 176 spectra in this category, and most of them are located in the vicinity of the bright filaments or diffuse interior (see the spatial distribution of black circles in Figure \ref{fig:dss2}). This suggests Group III spectra are likely to be associated with the SNR, and further investigation will clarify their origin.

Group IV is for stellar-emission only spectra. 234 spectra are classified based on no evidence for any association to the SNR. Figure \ref{fig:dss2} shows that Group IV spectra (yellow circles) are uniformly distributed over the entire remnant in general, which supports their non-association with the SNR. 

\begin{deluxetable}{cccc}[hbt!]
\tablecaption{Spectrum Classification for 778 Spectra after the First Screening 
\label{tab:specgr}}
\tablewidth{0pt}
\tablehead{
\colhead{Group} & \colhead{Number} & \colhead{Note} & \colhead{Symbol Color\tablenotemark{a}}
}
\startdata
I & 75 & SNR-dominated & red \\
IIa & 79 & (strong) SNR + stellar & cyan \\
IIb & 214 & (weak) SNR + (strong) stellar & blue \\
III & 176 & Ambiguous, possibly Balmer-dominated & black \\
IV & 234 & Stellar-dominated & yellow\\
\enddata
\tablecomments{Group I and IIa are mostly used for analysis in Section \ref{sec:res}, and IIb is only used to estimate $n_{\rm e}$ (see Section \ref{subsec:Te_ne}).}
\tablenotetext{a}{Symbol colors in Figure \ref{fig:dss2}}
\end{deluxetable}

\clearpage

\subsection{Line Identification}\label{subsec:line}

For Group I and II, line identification is carried out to measure line intensities and to derive relative ratios. Those with high signal-to-noise ratio (S/N) clearly exhibit various emission lines as previously reported \citep[e.g.,][]{fesen82,fesen96}: [\ion{O}{2}] $\lambda$3727, [\ion{Ne}{3}] $\lambda$3869, [\ion{S}{2}] $\lambda\lambda$4069, 4076,  [\ion{O}{3}] $\lambda$4363, [\ion{Fe}{3}] $\lambda$4658, \ion{He}{2} $\lambda$4686, [\ion{O}{3}] $\lambda\lambda$4959, 5007, [\ion{N}{1}] $\lambda$5200, [\ion{N}{2}] $\lambda$5755, \ion{He}{1} $\lambda$5876, [\ion{O}{1}] $\lambda\lambda$6300, 6364, [\ion{N}{2}] $\lambda\lambda$6548, 6583, [\ion{S}{2}] $\lambda\lambda$6717, 6731, [\ion{Ca}{2}] $\lambda\lambda$7291, 7324, [\ion{O}{2}] $\lambda\lambda$7320, 7330, and the Balmer lines (e.g., H$\alpha$, H$\beta$, H$\gamma$, etc.)\footnote{%
    Unless otherwise specified, [\ion{Ne}{3}] $\lambda$3869, [\ion{O}{3}] $\lambda\lambda$4959, 5007, [\ion{N}{1}] $\lambda$5200, [\ion{O}{1}] $\lambda\lambda$6300, 6364, [\ion{N}{2}] $\lambda\lambda$6548, 6583, and [\ion{S}{2}] $\lambda\lambda$6717, 6731 are shortly [\ion{Ne}{3}], [\ion{O}{3}] $\lambda4959+$, [\ion{N}{1}], [\ion{O}{1}] $\lambda6300+$, [\ion{N}{2}] $\lambda6548+$, and [\ion{S}{2}] $\lambda6717+$, respectively.}. %
Various weak lines are also detected including [\ion{Fe}{2}] $\lambda$4359, [\ion{Fe}{3}] $\lambda$4986, [\ion{Fe}{2}] $\lambda$5158, [\ion{S}{3}] $\lambda$6312, \ion{He}{1} $\lambda$7065, [\ion{Ar}{3}] $\lambda$7136, [\ion{Fe}{2}] $\lambda$7155, \ion{He}{1} $\lambda$7281, [\ion{Ni}{2}] $\lambda$7378. To handle bulk data consistently and efficiently, we do not aim to fit every detected line but do those needed for further analysis. Consequently, intensities of 15 emission lines (i.e., [\ion{O}{2}] $\lambda3727$, [\ion{Ne}{3}], [\ion{O}{3}] $\lambda$4363, H$\beta$, [\ion{O}{3}] $\lambda$4959+, [\ion{N}{1}], [\ion{N}{2}] $\lambda$5755, [\ion{O}{1}] $\lambda6300+$, [\ion{N}{2}] $\lambda$6548+, H$\alpha$, [\ion{S}{2}] $\lambda$6717+) are obtained for Group I and IIa, and only [\ion{S}{2}] $\lambda$6717+ are obtained for Group IIb. 

For Group I spectra (i.e., SNR-dominated spectra), line intensities are measured by a Gaussian fit with a linear baseline to each line profile. When two or more emission lines are adjacent such as H$\alpha$ and [\ion{N}{2}] $\lambda6548+$ or [\ion{S}{2}] $\lambda6717+$, a single baseline from a wider wavelength range is used for all these lines. Integrated intensities normalized to H$\beta$ are listed in Table \ref{tab:gr1}. Although LAMOST sky subtraction using principal component analysis reduces the averages of residuals down to $\sim3\%$ \citep{bai17}, several emission lines, especially weak lines, could still be contaminated by the residuals. For instance, relatively weak [\ion{O}{3}] $\lambda$4363 might be affected by Hg line at 4358 \AA~arising from mercury streetlights, and [\ion{O}{1}] $\lambda6300+$ can be contaminated by imperfect subtraction of the strong [\ion{O}{1}] night sky emission. When the determination of the baseline is problematic or the emission feature is damaged, only upper limits are quoted.

\begin{figure}
    \epsscale{0.7}
    \plotone{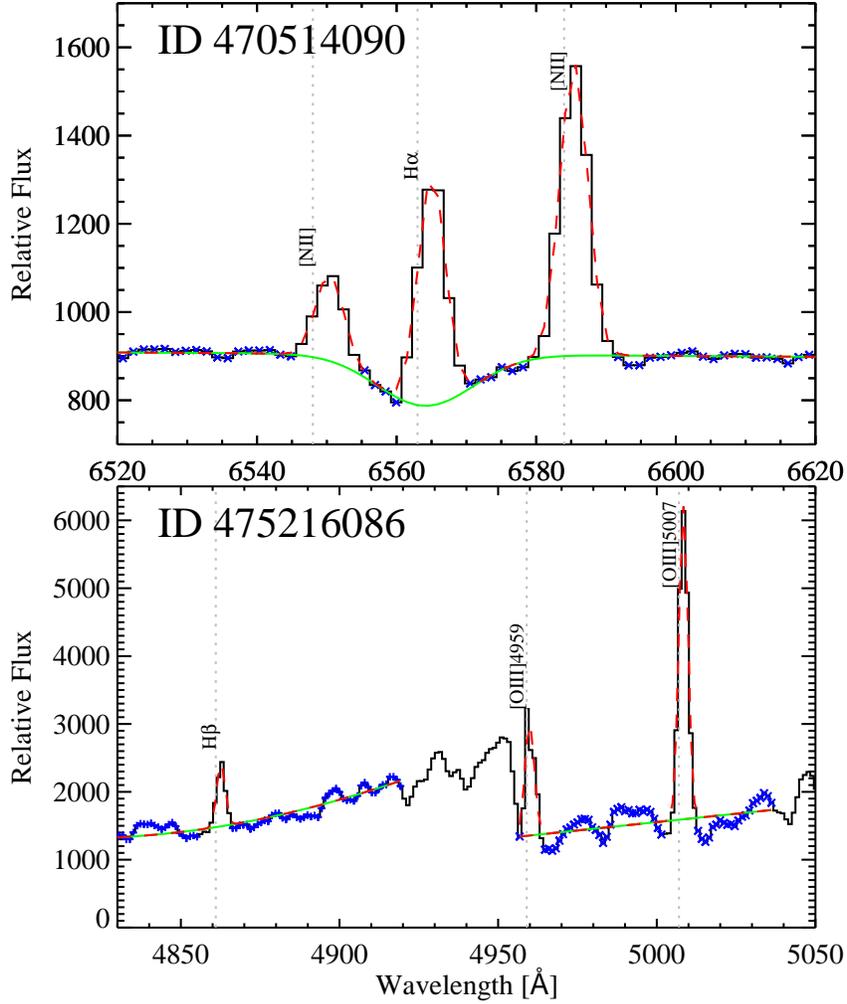}
    \caption{Example line fitting results for Group IIa spectra that require additional treatment (see details in Section \ref{subsec:line}). Obs. ID 470514090 spectrum near H$\alpha$ and obs. ID 475216086 near H$\beta$ (both also shown in Figure \ref{fig:EXspe}) are presented in upper and lower panels, respectively. Each plot shows the spectrum in black, the full best-fit model in red, and the baseline in green. Data points used for baseline fit are marked with blue ``x'' symbols. For obs. ID 475216086, [\ion{O}{3}] $\lambda$4959, 5007 share their baseline, which is determined with a linear fitting around the selected wavelength range. Because the continuum near [\ion{O}{3}] $\lambda$4959 is affected by TiO bands, the baseline only includes its redward range. 
    }
    \label{fig:c1fit}
\end{figure}

For Group IIa spectra (i.e., SNR+stellar spectra), most of line intensities are measured in the same way as Group I. When hydrogen absorption is significant or other stellar features contaminate neighborhood of an emission line, however, additional treatment is applied (see examples in Figure \ref{fig:c1fit}). For H$\alpha$ or H$\beta$ lying on top of a stellar absorption feature, the absorption feature is first fitted with a negative Gaussian (or a Lorentz profile for some cases with a wider wing). For instance, Obs. ID 470514090 shows a series of H absorption features, which is fitted by a negative gaussian (green line in Figure \ref{fig:c1fit}). Then, the fitted absorption profile is used as a baseline to fit the emission line from the SNR. Also, when the star along the line of sight is M-type, TiO bands can dominate its spectrum. In such cases, [\ion{O}{3}] $\lambda$4959 is adjacent to one of TiO bands (e.g., see obs. ID 475216086 in Figure \ref{fig:c1fit}), so we only use its redward range (marked with blue ``x'' in Figure \ref{fig:c1fit}) for baseline fitting. Measured line intensities relative to H$\beta$ for Group IIa spectra are listed in Table \ref{tab:gr2a}.

In spite of $10\%$ precision expected for high S/N spectra \citep[see also Section \ref{subsec:lamost}]{xiang15}, note that there are indications of larger errors in some spectra. The Balmer decrements, H$\alpha$/H$\beta$, should be about close to 2.9 in the recombining plasma of SNR shocks \citep{hummerstorey}, and the typical reddening to the Cygnus Loop E(B-V) \citep{fesen18} would increase that ratio to 3.1-3.2. Spectra in Tables \ref{tab:gr1}--\ref{tab:gr2a} below span the range from 2.02 to 6.25. Relatively slow shocks in partly neutral gas can produce higher Balmer decrements \citep{raymond79}, but the high [O~III]/H$\beta$ ratios of those spectra show that they are not such slow shocks. A higher reddening could account for some of the spectra with large Balmer decrements, and the western limb of the Cygnus Loop shows the interaction of the shock with a dense cloud having E(B-V) up to about 0.5 \citep{fesen18}, but even that would not account for Balmer decrements above 4. We conclude that some of the measured Balmer line ratios are in error by factors of 1.5 or more. This might be an offset between the red and blue sections of the spectra, but there is no obvious correlation between Balmer decrements and ratios between other lines at the red and blue ends of the spectrum. We therefore caution that the uncertainties are larger than might have been expected, and those with H$\alpha$/H$\beta<$2.9 or $>4.0$ are denoted with open symbols in Figures \ref{fig:corr_elm}--\ref{fig:Te_N2}.

\startlongtable
\begin{deluxetable}{crrrrrrrrrrrrrrr@{$\pm$}l}
\tabletypesize{\scriptsize}

\tablecaption{Relative Line Intensities for Group I Spectra relative to H$\beta$ (H$\beta$=100)\label{tab:gr1}}


\tablehead{
\colhead{Obs. ID} & \colhead{[\ion{O}{2}]} & \colhead{[\ion{Ne}{3}]} & \colhead{[\ion{O}{3}]}  & \colhead{[\ion{O}{3}]} & \colhead{[\ion{O}{3}]} & \colhead{[\ion{N}{1}]} & \colhead{[\ion{N}{2}]} & \colhead{[\ion{O}{1}]} & \colhead{[\ion{O}{1}]} & \colhead{[\ion{N}{2}]} & \colhead{H$\alpha$} & \colhead{[\ion{N}{2}]} & \colhead{[\ion{S}{2}]} & \colhead{[\ion{S}{2}]} & \multicolumn{2}{c}{H$\beta$ Flux\tablenotemark{a}} \\ 
\colhead{} & \colhead{$\lambda$3727} & \colhead{$\lambda$3869} & \colhead{$\lambda$4363} & \colhead{$\lambda$4959} & \colhead{$\lambda$5007} & \colhead{$\lambda$5200} & \colhead{$\lambda$5755} & \colhead{$\lambda$6300} & \colhead{$\lambda$6364} & \colhead{$\lambda$6548} & \colhead{$\lambda$6564} & \colhead{$\lambda$6583} & \colhead{$\lambda$6717} & \colhead{$\lambda$6731} & \multicolumn{2}{c}{(counts)} } 

\startdata
 470503073 &  2550 &   158 &  \nodata &   142 &   486 &  \nodata &  \nodata &    84 &  \nodata &   142 &   272 &   462 &   239 &   191 &   306 &   58 \\
 470503103 &  2725 &   137 &    63 &   202 &   650 &  \nodata &  \nodata &    76 &    24 &   146 &   483 &   441 &   234 &   213 &   220 &   37 \\
 470503109 &  1212 &    64 &    22 &    59 &   175 &    8 &    5 &    55 &    15 &    52 &   275 &   167 &   264 &   240 &  1833 &   189 \\
 470503144 &   785 &    34 &  \nodata &    26 &    95 &    14 &  \nodata &    64 &    18 &    69 &   318 &   222 &   121 &    97 &   734 &   84 \\
 470503149 &   707 &  \nodata &  \nodata &    24 &  \nodata &    29 &  \nodata &   251 &    59 &    63 &   406 &   213 &    79 &    60 &   154 &   20 \\
 470503216 &  2260 &   214 &    84 &   342 &  1077 &  \nodata &    11 &    18 &    3 &   103 &   290 &   325 &   192 &   139 &  1297 &   143 \\
 470504035 &  1338 &    83 &    34 &   110 &   321 &    15 &  \nodata &    69 &    20 &    70 &   301 &   240 &   262 &   190 &   172 &    19 \\
 470504061 &  1056 &    89 &    37 &   126 &   394 &  \nodata &    7 &    16 &    6 &    28 &   264 &    98 &    61 &    45 &   409 &   43 \\
 470504132 &  1026 &  \nodata &  \nodata &  \nodata &    34 &    31 &  \nodata &   142 &    48 &    66 &   237 &   202 &   178 &   130 &   345 &    36 \\
 470504139 &  1314 &   165 &    78 &   313 &   936 &  \nodata &  \nodata &    45 &    16 &    36 &   345 &   126 &   116 &    86 &   202 &    22 \\
 470504144 &  1125 &    63 &    18 &    76 &   213 &    4 &    5 &    36 &    12 &    59 &   305 &   180 &   117 &    85 &   878 &    88 \\
 470504151 &  1105 &  \nodata &  \nodata &    13 &    30 &    32 &  \nodata &   150 &    51 &    67 &   306 &   235 &   267 &   203 &   272 &    29 \\
 470505053 &   726 &  \nodata &  \nodata &    68 &   239 &    27 &  \nodata &   108 &    49 &    72 &   440 &   258 &   181 &   136 &   166 &    19 \\
 470509027 &  2123 &   245 &   127 &   445 &  1377 &  \nodata &  \nodata &    29 &    5 &    79 &   320 &   281 &   192 &   153 &   940 &   101 \\
 470509075 &  1275 &    79 &  \nodata &    64 &   221 &  \nodata &  \nodata &    56 &    14 &    81 &   299 &   230 &   188 &   159 &   311 &   34 \\
 470509080 &   556 &    13 &  \nodata &    9 &    28 &    34 &  \nodata &   115 &    36 &    58 &   328 &   196 &   144 &   119 &  1462 &   149 \\
 470509089 &  2170 &   178 &    90 &   284 &   777 &  \nodata &  \nodata &   111 &    23 &   119 &   337 &   336 &   175 &   147 &   251 &   30 \\
 470509097 &  1129 &    55 &    23 &    64 &   218 &    16 &  \nodata &    61 &    14 &    69 &   268 &   222 &   160 &   125 &  2344 &   240 \\
 470509098 &   861 &   270 &   162 &   535 &  1558 &  \nodata &  \nodata &   105 &    28 &    40 &   397 &    99 &    44 &    34 &   117 &   15 \\
 470511036 &  2296 &   156 &    54 &   153 &   488 &    34 &    9 &   153 &    51 &   161 &   514 &   567 &   367 &   276 &  1411 &  163 \\
 470511039 &   786 &    19 &  \nodata &    17 &    60 &    16 &  \nodata &    85 &    26 &    63 &   282 &   197 &   174 &   132 &  1149 &  117 \\
 470511109 &   790 &    22 &    3 &    13 &    41 &    31 &    5 &   124 &    40 &    79 &   346 &   216 &   188 &   138 &  2031 &   203 \\
 470511160 &  1048 &    39 &    9 &    32 &   108 &    14 &    5 &    73 &    17 &    74 &   285 &   268 &   193 &   154 &  1341 &  141 \\
 470511161 &   926 &    40 &  \nodata &    46 &   160 &    18 &    6 &    92 &    14 &    74 &   286 &   250 &   180 &   136 &   758 &  81 \\
 470511167 &   762 &    38 &  \nodata &    36 &   116 &    14 &  \nodata &    56 &    13 &    87 &   345 &   274 &   217 &   161 &   642 &   66 \\
 470511185 &  1725 &  \nodata &  \nodata &    58 &   221 &  \nodata &  \nodata &   220 &    54 &   138 &   359 &   445 &   262 &   198 &    78 &    13 \\
 470511201 &   845 &  \nodata &  \nodata &    11 &    36 &    27 &  \nodata &   195 &    51 &   111 &   344 &   375 &   135 &   104 &   263 &   29 \\
 470511223 &  1700 &  \nodata &  \nodata &   132 &   307 &  \nodata &  \nodata &   155 &    74 &   124 &   292 &   350 &   177 &   133 &    97 &   14 \\
 470512067 &  1116 &    64 &  \nodata &    54 &   153 &  \nodata &  \nodata &    52 &    16 &    74 &   265 &   224 &   143 &   108 &   191 &   21 \\
 470512069 &   868 &    48 &    13 &    45 &   137 &    14 &    4 &    62 &    17 &    70 &   268 &   217 &   172 &   134 &  1450 &   146 \\
 470512080 &  1001 &    81 &    31 &   106 &   319 &    18 &    5 &    77 &    25 &    73 &   279 &   243 &   182 &   139 &  1252 &   126 \\
 470512083 &  1029 &   124 &    42 &   166 &   482 &  \nodata &  \nodata &    7 &    7 &    78 &   460 &   284 &   194 &   154 &   162 &    19 \\
 470512084 &  1808 &   179 &    60 &   208 &   612 &    12 &    10 &    82 &    22 &   109 &   344 &   355 &   245 &   203 &  1899 &  219 \\
 470512087 &  1406 &   152 &    54 &   240 &   735 &    5 &    6 &    33 &    9 &    71 &   263 &   209 &   118 &    93 &  1002 &   100 \\
 470512129 &  1585 &   197 &    70 &   227 &   670 &  \nodata &  \nodata &    34 &    9 &    70 &   245 &   206 &   122 &    98 &   195 &  21 \\
 470512135 &   798 &  \nodata &  \nodata &   114 &   278 &  \nodata &  \nodata &    30 &    12 &    20 &   202 &    70 &    44 &    42 &    52 &    7 \\
 470512137 &   438 &    16 &  \nodata &    3 &    11 &    31 &    3 &    63 &    20 &    33 &   179 &   103 &   129 &    90 &  1585 &   159 \\
 470512241 &  1263 &   186 &    75 &   291 &   884 &    9 &    6 &    25 &    2 &    53 &   302 &   192 &   146 &   116 &   881 &   92 \\
 470514093 &  1778 &   128 &    39 &   190 &   561 &    18 &    6 &    79 &    15 &   112 &   309 &   352 &   237 &   177 &   888 &   97 \\
 470514094 &  1563 &   100 &    41 &   116 &   369 &    19 &    4 &    85 &    21 &    99 &   305 &   303 &   157 &   119 &  1048 &  110 \\
 470514096 &  1122 &   107 &    42 &   184 &   549 &  \nodata &    5 &    19 &    6 &    52 &   229 &   168 &   111 &   120 &   425 &   43 \\
 470514141 &  1140 &  \nodata &    72 &   210 &   632 &    21 &  \nodata &    44 &    10 &    50 &   309 &   197 &   121 &    79 &   148 &  17 \\
 470514142 &  1632 &   120 &    36 &   146 &   445 &    11 &    6 &    45 &    12 &    78 &   287 &   298 &   144 &   106 &  2024 &   205 \\
 470514165 &  2973 &   377 &   174 &   642 &  1910 &  \nodata &    11 &    23 &    3 &    59 &   324 &   247 &   123 &    97 &   741 &  77 \\
 470514168 &  1379 &    68 &    19 &    81 &   233 &    12 &    3 &    52 &    13 &    65 &   282 &   258 &   143 &   107 &  1159 &   116 \\
 470515006 &  2625 &   158 &    76 &   243 &   753 &  \nodata &  \nodata &    31 &  \nodata &    91 &   257 &   246 &   198 &   194 &   228 &   42 \\
 470515065 &  3939 &  \nodata &  \nodata &   146 &   473 &  \nodata &  \nodata &   120 &    65 &   194 &   531 &   679 &   277 &   206 &    48 &   10 \\
 470515199 &  2253 &   148 &    80 &   210 &   709 &  \nodata &  \nodata &  \nodata &  \nodata &    49 &   425 &   188 &   105 &    92 &   113 &   19 \\
 470516089 &   948 &    38 &    14 &    44 &   125 &    40 &  \nodata &   128 &    36 &    69 &   271 &   232 &   196 &   153 &  1876 &   195 \\
 470516108 &  \nodata &  \nodata &    85 &   192 &   586 &  \nodata &  \nodata &    50 &  \nodata &  \nodata &   343 &  \nodata &    19 &  \nodata &    46 & 10 \\
 470516172 &   811 &  \nodata &  \nodata &  \nodata &    44 &    8 &  \nodata &    33 &    20 &    60 &   320 &   202 &   185 &   125 &   300 &    31 \\
 470516174 &  1337 &    64 &    12 &    49 &   159 &    20 &    8 &    85 &    24 &    84 &   322 &   269 &   191 &   143 &   630 &   66 \\
 470516184 &  1813 &  \nodata &  \nodata &  \nodata &   296 &  \nodata &  \nodata &    87 &  \nodata &    83 &   404 &   314 &   389 &   281 &    49 &   11 \\
 470516229 &  1271 &    74 &    16 &    78 &   238 &    14 &    5 &    69 &    18 &    67 &   280 &   257 &   209 &   175 &  1131 &   114 \\
 470516246 &  1740 &   126 &    34 &   169 &   516 &    8 &    7 &    30 &    8 &    64 &   226 &   231 &   111 &    83 &  1271 &   128 \\
 475203151 &  2148 &    91 &  \nodata &    91 &   336 &   135 &  \nodata &   336 &    69 &   270 &   517 &   878 &   367 &   283 &  1065 &  164 \\
 475211039 &   948 &    43 &  \nodata &    51 &   149 &    11 &  \nodata &   139 &    43 &   118 &   625 &   470 &   230 &   244 &  1558 &  166  \\
 475211106 &   562 &    19 &    2 &    13 &    45 &    14 &    3 &    86 &    27 &    63 &   331 &   294 &   100 &   322 &  5080 &   510 \\
 475211136 &  1667 &    76 &    41 &   216 &   282 &    21 &  \nodata &   192 &    48 &   160 &   519 &   492 &   214 &   167 &  1203 &   131 \\
 475211152 &  1360 &    67 &    15 &    96 &   313 &  \nodata &    9 &    24 &    12 &   110 &   412 &   366 &   291 &   224 &  1275 &   140 \\
 475211160 &   954 &    19 &  \nodata &    43 &   146 &    28 &  \nodata &   164 &    57 &   128 &   502 &   420 &   296 &   236 &  1065 &  119 \\
 475211213 &  1432 &  \nodata &  \nodata &  \nodata &   171 &  \nodata &  \nodata &    97 &    28 &   103 &   440 &   310 &   279 &   201 &   214 &   50 \\
 475212060 &  1106 &  \nodata &  \nodata &    86 &   206 &  \nodata &  \nodata &    26 &    10 &    58 &   210 &   182 &    64 &    55 &   125 &   24 \\
 475212080 &  1651 &   150 &    53 &   296 &   919 &    10 &    6 &    38 &    8 &   104 &   305 &   322 &   247 &   203 &  5351 &  555 \\
 475212129 &   789 &    31 &  \nodata &    41 &   130 &  \nodata &  \nodata &    38 &    5 &    48 &   252 &   153 &   162 &   122 &   576 &   60 \\
 475212236 &  1318 &    65 &    50 &   102 &   306 &    45 &  \nodata &   124 &    18 &   121 &   431 &   355 &   276 &   237 &  3264 &  494 \\
 475214088 &  1502 &   117 &    56 &   180 &   727 &  \nodata &  \nodata &    50 &  \nodata &   143 &   354 &   694 &   340 &   243 &  1800 &  247 \\
 475214135 &   683 &  \nodata &  \nodata &    16 &    42 &    29 &  \nodata &   252 &    75 &   146 &   365 &   342 &   246 &   169 &   977 &   100 \\
 475215159 &   673 &    25 &    8 &    50 &   177 &    24 &  \nodata &   138 &    38 &    81 &   419 &   254 &   132 &    97 &   781 &   82 \\
 475215190 &  1632 &   235 &   109 &   526 &  1674 &  \nodata &  \nodata &   279 &  \nodata &    37 &   545 &   165 &    60 &  \nodata &    66 &   18 \\
 475216143 &   862 &    46 &    8 &    62 &   211 &    30 &    8 &   103 &    34 &    95 &   300 &   309 &   235 &   161 &  2308 &   233 \\
 475216147 &   717 &    37 &    7 &    29 &    97 &    55 &    7 &   139 &    44 &    81 &   374 &   257 &   228 &   172 &  3106 &   49 \\
 475216228 &  2404 &   179 &  \nodata &   112 &   408 &  \nodata &  \nodata &   144 &    49 &   196 &   440 &   598 &   269 &   214 &   301 &   47 \\
 475216229 &  1236 &   155 &    32 &   110 &   383 &  \nodata &  \nodata &   226 &    68 &   159 &   425 &   441 &   331 &   254 &   393 &   49 \\
 475216241 &   708 &    33 &  \nodata &  \nodata &    39 &    16 &  \nodata &   123 &    40 &   117 &   429 &   341 &   239 &   177 &   976 &   105 \\
\enddata


\tablecomments{\nodata used for non-detection}
\tablenotetext{a}{Note that H$\beta$ flux quoted here does not necessarily represent the absolute brightness at the position of a given fiber since the LAMOST spectra are not absolute-calibrated (see Section \ref{subsec:lamost}). However, it is worthwhile to list them because the flux with uncertainty gives an idea for the S/N of the entire spectrum to some extent. The final uncertainty is the quadratic sum of the calibration errors (10\% adopted) and the flux uncertainties mainly arising fluctuation of the baseline during a linear fitting (see Section \ref{subsec:line}).}

\end{deluxetable}
\startlongtable
\begin{deluxetable}{crrrrrrrrrrrrrrr@{$\pm$}l}
\tabletypesize{\scriptsize}

\tablecaption{Relative Line Intensities for Group IIa Spectra relative to H$\beta$ (H$\beta$=100)\label{tab:gr2a}}


\tablehead{
\colhead{Obs. ID} & \colhead{[\ion{O}{2}]} & \colhead{[\ion{Ne}{3}]} & \colhead{[\ion{O}{3}]}  & \colhead{[\ion{O}{3}]} & \colhead{[\ion{O}{3}]} & \colhead{[\ion{N}{1}]} & \colhead{[\ion{N}{2}]} & \colhead{[\ion{O}{1}]} & \colhead{[\ion{O}{1}]} & \colhead{[\ion{N}{2}]} & \colhead{H$\alpha$} & \colhead{[\ion{N}{2}]} & \colhead{[\ion{S}{2}]} & \colhead{[\ion{S}{2}]} & \multicolumn{2}{c}{H$\beta$ flux\tablenotemark{a}} \\ 
\colhead{} & \colhead{$\lambda$3727} & \colhead{$\lambda$3869} & \colhead{$\lambda$4363} & \colhead{$\lambda$4959} & \colhead{$\lambda$5007} & \colhead{$\lambda$5200} & \colhead{$\lambda$5755} & \colhead{$\lambda$6300} & \colhead{$\lambda$6364} & \colhead{$\lambda$6548} & \colhead{$\lambda$6564} & \colhead{$\lambda$6583} & \colhead{$\lambda$6717} & \colhead{$\lambda$6731} & \multicolumn{2}{c}{(counts)} }
\startdata
 470503207 &    927 &   \nodata &   \nodata &   \nodata &    101 &   \nodata &   \nodata &     41 &   \nodata &     65 &    507 &    190 &     42 &     59 &    500 &    110 \\
 470504019 &   1617 &   \nodata &   \nodata &     27 &     88 &     72 &   \nodata &    175 &     25 &    172 &    344 &    571 &    242 &    212 &    411 &    60 \\
 470504066 &   1979 &    120 &     53 &    104 &    337 &   \nodata &      8 &      6 &   \nodata &     51 &    239 &    184 &    107 &     91 &    281 &    39 \\
 470509065 &   1306 &     72 &   \nodata &   \nodata &     92 &   \nodata &   \nodata &     90 &   \nodata &     90 &    337 &    320 &    260 &    222 &    293 &    56 \\
 470511206 &   1196 &   \nodata &   \nodata &   \nodata &    279 &   \nodata &   \nodata &     95 &   \nodata &     45 &    235 &    186 &     98 &     95 &     26 &    11 \\
 470512054 &   \nodata &   \nodata &   \nodata &     46 &    123 &   \nodata &   \nodata &     35 &      7 &      6 &    281 &     39 &   \nodata &   \nodata &    124 &    16 \\
 470512145 &   \nodata &   \nodata &   \nodata &     42 &    105 &   \nodata &   \nodata &     91 &     34 &     20 &    160 &     55 &     30 &     42 &     27 &     6 \\
 470514079 &   3450 &    110 &   \nodata &    119 &    380 &   \nodata &   \nodata &    338 &   \nodata &    328 &    496 &   1011 &    381 &    360 &    249 &   102 \\
 470514157 &    410 &     47 &     28 &     67 &    208 &   \nodata &   \nodata &     12 &      4 &      9 &    308 &     42 &     28 &     28 &    433 &    51 \\
 470515071 &    972 &   \nodata &   \nodata &     56 &    252 &   \nodata &   \nodata &     64 &   \nodata &     65 &    453 &    217 &    125 &    115 &     88 &    23 \\
 470515072 &   1360 &     87 &   \nodata &   \nodata &   \nodata &     40 &   \nodata &     94 &   \nodata &    120 &    288 &    319 &    246 &    196 &     98 &    14 \\
 470515095 &   1051 &   \nodata &   \nodata &   \nodata &     40 &   \nodata &   \nodata &     37 &   \nodata &     62 &    266 &    200 &    151 &    127 &    118 &    18 \\
 470516241 &   2274 &     82 &   \nodata &     93 &    228 &   \nodata &   \nodata &     72 &   \nodata &    148 &    325 &    434 &    283 &    236 &    739 &   146 \\
 475209098 &    530 &   \nodata &   \nodata &     55 &    176 &   \nodata &   \nodata &     38 &      9 &     28 &    197 &     99 &     72 &     55 &    305 &    40 \\
 475211030 &   1372 &     40 &   \nodata &     61 &    230 &   \nodata &   \nodata &     63 &   \nodata &    135 &    699 &    432 &    278 &    252 &   2009 &   394 \\
 475211036 &   3571 &    150 &   \nodata &    273 &    835 &   \nodata &   \nodata &    240 &   \nodata &    304 &    794 &   1031 &    756 &    708 &   1138 &   503 \\
 475214165 &   4116 &    435 &    272 &   1069 &   2288 &   \nodata &   \nodata &   \nodata &   \nodata &    301 &    382 &   1012 &    359 &    577 &    762 &   286 \\
 470503118 &   2992 &    206 &     62 &    283 &    852 &   \nodata &   \nodata &     48 &     17 &    106 &    322 &    338 &    292 &    214 &    444 &    46 \\
 470503142 &   1294 &     91 &     55 &    133 &    379 &   \nodata &     13 &    103 &   \nodata &    108 &    319 &    333 &    153 &    160 &    284 &    44 \\
 470503172 &   1622 &     62 &   \nodata &     54 &    264 &   \nodata &   \nodata &     62 &   \nodata &     85 &    312 &    308 &    196 &    175 &    433 &    71 \\
 470503215 &    922 &     54 &   \nodata &   \nodata &     82 &   \nodata &   \nodata &    176 &     43 &    124 &    326 &    358 &    173 &    124 &    313 &    73 \\
 470504022 &    845 &   \nodata &   \nodata &   \nodata &     92 &     79 &   \nodata &    230 &     78 &    160 &    244 &    443 &    112 &     82 &     92 &    19 \\
 470504194 &   1085 &   \nodata &   \nodata &     58 &    203 &   \nodata &   \nodata &   \nodata &   \nodata &     40 &    515 &    173 &     46 &     98 &    467 &   135 \\
 470505067 &   2997 &    131 &    116 &    216 &    748 &     79 &   \nodata &    149 &   \nodata &    227 &    552 &    630 &    537 &    455 &    378 &   112 \\
 470509055 &   1847 &    174 &   \nodata &    145 &    474 &   \nodata &   \nodata &    104 &     24 &     78 &    281 &    242 &    195 &    155 &    141 &    18 \\
 470509084 &   1185 &     35 &   \nodata &     88 &    255 &     18 &     15 &     34 &   \nodata &     63 &    240 &    184 &    183 &    163 &    222 &    36 \\
 470511031 &   1699 &    130 &     47 &    154 &    469 &     24 &      8 &     80 &     26 &     91 &    361 &    291 &    159 &    118 &   1870 &    189 \\
 470511035 &    743 &     54 &   \nodata &     13 &     77 &   \nodata &   \nodata &     72 &     20 &     47 &    440 &    161 &     30 &     28 &    339 &    37 \\
 470511037 &   1317 &     57 &   \nodata &     25 &    103 &     26 &   \nodata &    126 &     39 &    105 &    309 &    323 &    189 &    146 &   1594 &   161 \\
 470511124 &   2027 &   \nodata &   \nodata &     26 &    125 &   \nodata &   \nodata &    233 &     54 &    152 &    506 &    532 &    324 &    245 &     79 &    19 \\
 470511144 &   1395 &   \nodata &   \nodata &   \nodata &   \nodata &     98 &   \nodata &    363 &    120 &    180 &    314 &    548 &    233 &    176 &    170 &    23 \\
 470511154 &    908 &   \nodata &   \nodata &   \nodata &     92 &   \nodata &   \nodata &     74 &     44 &     67 &    442 &    275 &    228 &    179 &    111 &    22 \\
 470511157 &   2648 &   \nodata &   \nodata &   \nodata &    365 &   \nodata &   \nodata &    200 &     35 &    205 &    757 &    766 &    575 &    475 &    108 &    23 \\
 470512051 &    819 &     64 &     21 &     55 &    172 &     31 &      3 &    103 &     34 &     77 &    333 &    240 &    197 &    152 &   1484 &   150 \\
 470512052 &    963 &     59 &     15 &     40 &    115 &     31 &      3 &    112 &     37 &     65 &    309 &    249 &    232 &    176 &    949 &    96 \\
 470512058 &    964 &     73 &     19 &     51 &    150 &     34 &      5 &    123 &     40 &     97 &    326 &    298 &    250 &    182 &    906 &    91 \\
 470512064 &    978 &    231 &   \nodata &    221 &    929 &   \nodata &   \nodata &     81 &   \nodata &     36 &    675 &    158 &     69 &     95 &     80 &    21 \\
 470512068 &    798 &    191 &    132 &    331 &    928 &   \nodata &   \nodata &    419 &   \nodata &     50 &    319 &    141 &     69 &     58 &    455 &    96 \\
 470512072 &   1434 &    167 &   \nodata &     56 &    190 &   \nodata &   \nodata &     90 &     22 &     93 &    280 &    314 &    229 &    178 &    333 &    65 \\
 470512073 &   1469 &     85 &   \nodata &     46 &    203 &   \nodata &   \nodata &     88 &      9 &    123 &    415 &    389 &    329 &    265 &    571 &    104 \\
 470512079 &   1571 &    171 &    102 &    207 &    591 &     30 &   \nodata &     77 &     22 &    151 &    450 &    397 &    335 &    286 &    249 &    45 \\
 470512099 &   1533 &   \nodata &   \nodata &    423 &   1237 &   \nodata &   \nodata &     55 &   \nodata &    108 &    787 &    420 &    180 &    220 &     20 &     7 \\
 470512147 &    846 &     91 &   \nodata &     59 &    180 &   \nodata &   \nodata &     35 &   \nodata &     67 &    315 &    211 &    259 &    205 &    353 &    46 \\
 470512236 &   1145 &   \nodata &   \nodata &     58 &    147 &   \nodata &   \nodata &     74 &   \nodata &     97 &    335 &    312 &    149 &    139 &    219 &    57 \\
 470514081 &    732 &   \nodata &   \nodata &     50 &     86 &   \nodata &   \nodata &    146 &     23 &    113 &    272 &    222 &    202 &    149 &     46 &     9 \\
 470514085 &   1382 &    124 &     58 &    225 &    656 &   \nodata &   \nodata &     35 &      7 &     76 &    310 &    268 &    233 &    171 &    252 &    28 \\
 470514087 &    980 &   \nodata &   \nodata &     31 &    100 &   \nodata &    128 &   \nodata &   \nodata &    167 &    251 &    298 &   \nodata &   \nodata &    213 &    35 \\
 470514089 &   1461 &     91 &     30 &    123 &    291 &   \nodata &     15 &   \nodata &   \nodata &    102 &    400 &    347 &    137 &    153 &    602 &    73 \\
 470514090 &   2655 &    271 &    118 &    166 &    497 &     42 &   \nodata &    164 &     51 &    176 &    456 &    588 &    319 &    255 &    567 &    89 \\
 470515085 &   1067 &   \nodata &   \nodata &     53 &    166 &   \nodata &     17 &    170 &     39 &     66 &    478 &    227 &    177 &    133 &    115 &    18 \\
 470515187 &   2537 &   \nodata &   \nodata &     57 &    215 &   \nodata &   \nodata &    188 &     53 &     71 &    357 &    312 &    261 &    192 &     27 &     7 \\
 470515213 &   1759 &   \nodata &   \nodata &     63 &    254 &   \nodata &   \nodata &   \nodata &   \nodata &     70 &    155 &    236 &    140 &    121 &    201 &    28 \\
 470516034 &   2530 &    325 &    281 &    553 &   1340 &   \nodata &   \nodata &     21 &   \nodata &     52 &    244 &    217 &    173 &    123 &     66 &     8 \\
 470516054 &   2414 &     61 &     50 &     72 &    219 &     23 &     11 &     68 &     17 &    123 &    361 &    389 &    267 &    203 &    687 &    87 \\
 470516057 &   1090 &     60 &     24 &     63 &    180 &     11 &      6 &     58 &     19 &     63 &    253 &    212 &    171 &    127 &    844 &    86 \\
 470516065 &    849 &   \nodata &   \nodata &   \nodata &   \nodata &     41 &   \nodata &    117 &     27 &     78 &    296 &    242 &    196 &    154 &    462 &    60 \\
 470516083 &   1551 &    129 &   \nodata &     88 &    337 &   \nodata &   \nodata &     93 &     19 &     68 &    454 &    227 &    152 &    130 &    347 &    57 \\
 470516084 &   2619 &    121 &   \nodata &    132 &    436 &   \nodata &   \nodata &     86 &   \nodata &    128 &    323 &    348 &    279 &    251 &    272 &    49 \\
 470516085 &   1738 &     88 &   \nodata &   \nodata &     45 &     27 &   \nodata &    108 &     38 &    121 &    343 &    403 &    211 &    167 &    180 &    24 \\
 470516086 &   2362 &    207 &     39 &    203 &    642 &   \nodata &   \nodata &     28 &      5 &    105 &    392 &    338 &    255 &    197 &    261 &    31 \\
 470516140 &   2215 &     89 &   \nodata &     86 &    241 &   \nodata &   \nodata &     85 &   \nodata &    137 &    285 &    419 &    170 &    136 &    522 &    74 \\
 470516144 &   2105 &    120 &   \nodata &     84 &    326 &   \nodata &   \nodata &     26 &   \nodata &    126 &    328 &    390 &    213 &    175 &    410 &    58 \\
 470516147 &    771 &     47 &   \nodata &     22 &     78 &     37 &      2 &    129 &     33 &     71 &    286 &    210 &    155 &    114 &    927 &    96 \\
 470516194 &   1938 &     86 &   \nodata &     26 &    119 &   \nodata &   \nodata &    107 &   \nodata &    153 &    432 &    446 &    294 &    271 &    293 &    52 \\
 470516231 &   1479 &   \nodata &   \nodata &     59 &    214 &     15 &     15 &    122 &     49 &    137 &    426 &    474 &    308 &    230 &    253 &    28 \\
 470516239 &   1658 &     96 &     49 &    179 &    622 &     26 &   \nodata &     30 &     17 &    107 &    401 &    397 &    216 &    171 &    256 &    38 \\
 470516248 &   3001 &    509 &    307 &    633 &   1879 &   \nodata &   \nodata &     37 &   \nodata &    107 &    725 &    413 &    177 &    155 &    156 &    29 \\
 475204144 &   1381 &   \nodata &   \nodata &     49 &    187 &     53 &      6 &    174 &     51 &    102 &    418 &    358 &    236 &    178 &    404 &    60 \\
 475211020 &    958 &   \nodata &   \nodata &   \nodata &    241 &   \nodata &   \nodata &   \nodata &     30 &    141 &    502 &    396 &    430 &    312 &    133 &    28 \\
 475211201 &    671 &   \nodata &   \nodata &     57 &    146 &   \nodata &   \nodata &     86 &     15 &     60 &    275 &    171 &    126 &    117 &    900 &   140 \\
 475211223 &   1328 &   \nodata &   \nodata &    130 &    316 &   \nodata &   \nodata &    153 &     41 &    108 &    306 &    286 &    156 &    123 &    238 &    71 \\
 475212052 &   2165 &   \nodata &   \nodata &    103 &    242 &   \nodata &   \nodata &    207 &   \nodata &    252 &    626 &    776 &    502 &    473 &    318 &   123 \\
 475212135 &   1283 &   \nodata &   \nodata &    145 &    465 &   \nodata &   \nodata &     45 &   \nodata &    123 &    477 &    389 &    387 &    322 &    172 &    39 \\
 475214211 &   1216 &   \nodata &     84 &    190 &    554 &     14 &   \nodata &    100 &     43 &    109 &    533 &    393 &    191 &    136 &    537 &    76 \\
 475215061 &   2714 &   \nodata &   \nodata &     42 &    139 &   \nodata &   \nodata &    150 &   \nodata &    243 &    330 &    723 &    371 &    338 &    272 &    83 \\
 475216063 &    776 &   \nodata &   \nodata &     68 &    229 &     81 &   \nodata &    191 &     56 &    100 &    443 &    349 &    358 &    292 &   2778 &   341 \\
 475216066 &   1654 &   \nodata &   \nodata &     45 &    214 &   \nodata &   \nodata &    198 &     69 &    182 &    556 &    626 &    671 &    381 &    361 &    50 \\
 475216086 &   1988 &    164 &     69 &    197 &    527 &   \nodata &    153 &   \nodata &   \nodata &    173 &    133 &    869 &    240 &    255 &   3126 &   511 \\
 475216246 &   1232 &     81 &      9 &     52 &    147 &     18 &      9 &    100 &     33 &     92 &    261 &    271 &    167 &    134 &   2173 &   226 \\
\enddata

\tablecomments{\nodata used for non-detection}
\tablenotetext{a}{Same as Table \ref{tab:gr1}.}


\end{deluxetable}

\section{Results} \label{sec:res}

\subsection{Variations and Correlations of Line Ratios} \label{subsec:ratio}

For the 154 Group I and IIa spectra in total, we measure line intensities relative to H$\beta$. To the best of our knowledge, this is the largest sample of optical spectra with line measurement observed in the Cygnus Loop, which does not just focus on bright filaments but cover the whole regions of the SNR. As listed in Table \ref{tab:gr1} and \ref{tab:gr2a} and presented in the sample spectra in Figure \ref{fig:EXspe}, the relative strengths of line emission vary significantly at different positions within the Cygnus Loop, and correlations among line ratios of different elements or different transitions are observed. As previously noticed \citep[e.g.,][hereafter, F82]{fesen82}, the intensity of [\ion{O}{3}] $\lambda4959+$ relative to H$\beta$ varies over two orders of magnitude (from 0.15 to $\ga25$ in Group I), and other lines such as [\ion{O}{2}] $\lambda3727$, [\ion{N}{2}] $\lambda6548+$, and [\ion{S}{2}] $\lambda6717+$ also vary in intensity over an order of magnitude, suggesting the presence of diverse physical conditions inside the single remnant. Note that dereddening is not applied here since the LAMOST spectra are only relatively flux-calibrated (Section \ref{subsec:lamost}). However, as extinction to the remnant is relatively low \citep[e.g., E(B-V)=0.08 mag,][]{parker67}, we postulate this does not affect our results significantly, in particular, when we compare emission lines nearby. To demonstrate the effect of extinction, line ratios using measured intensities and those of dereddened intensities in \citetalias{fesen82} are overplotted as a reference (see Figures \ref{fig:corr_elm}--\ref{fig:Te_N2}).

\subsubsection{Line ratios of different elements}

\begin{figure}[!ht]
\gridline{\fig{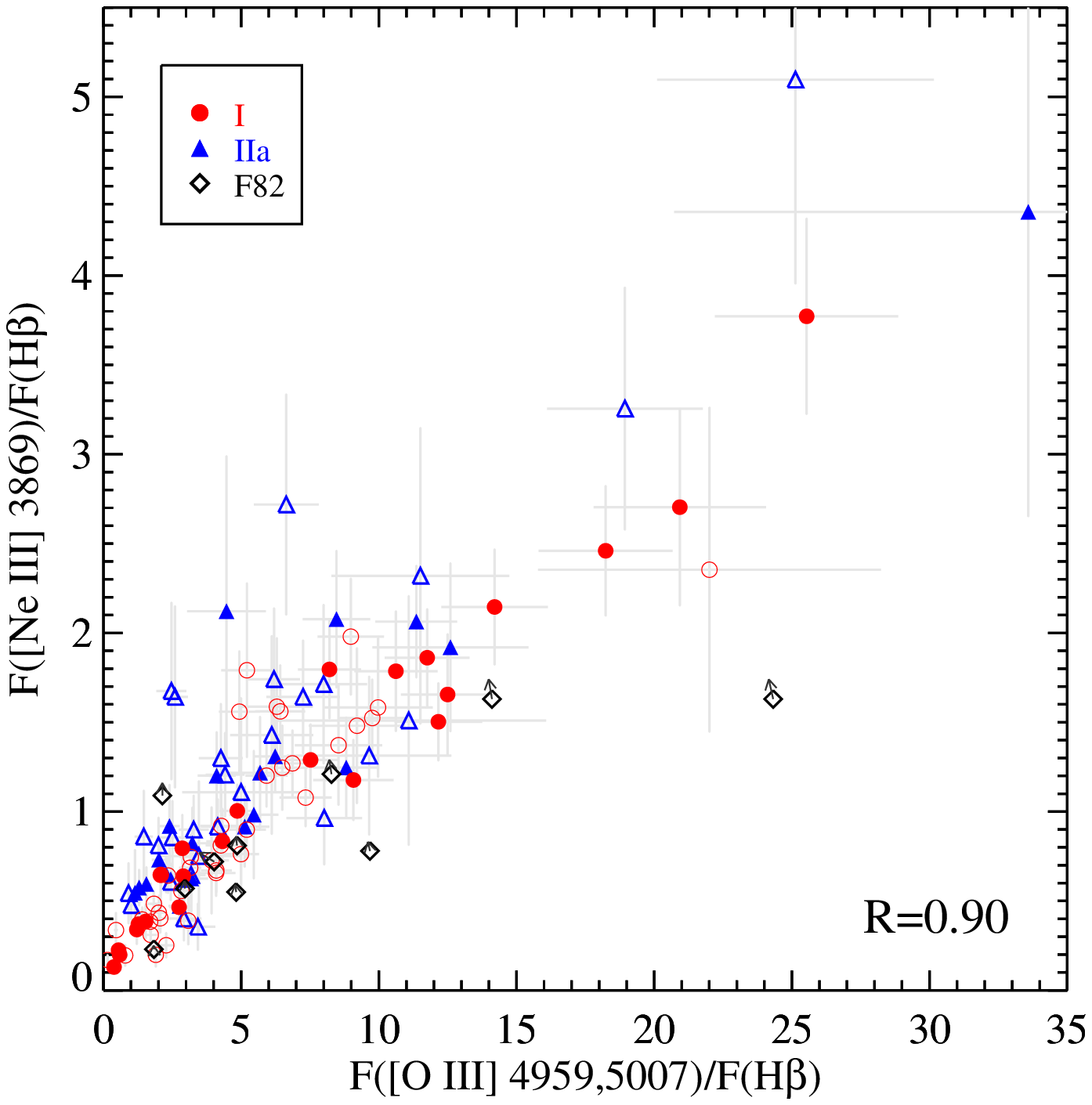}{0.5\textwidth}{a. [\ion{Ne}{3}] vs. [\ion{O}{3}] $\lambda4959+$}
          \fig{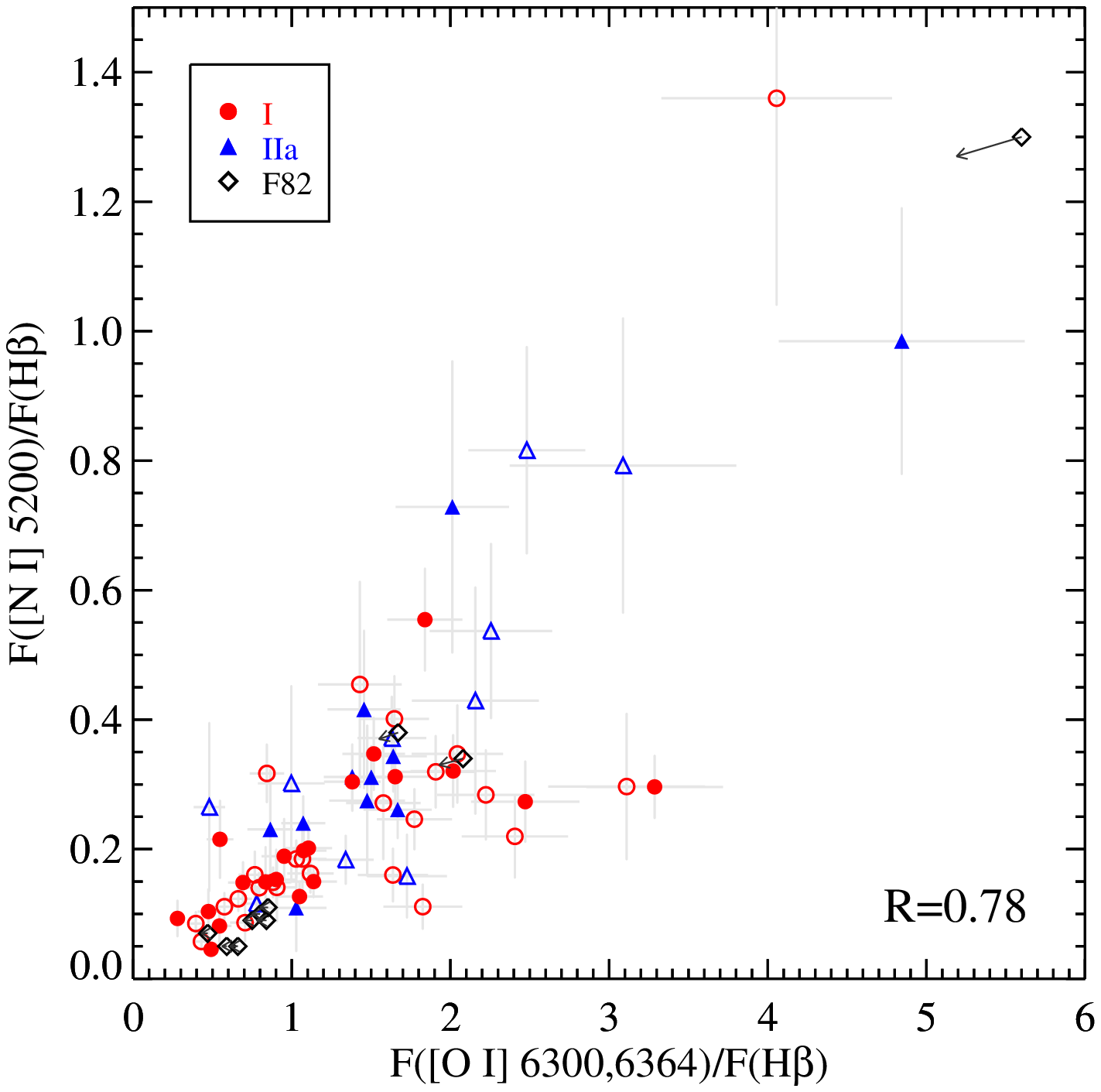}{0.5\textwidth}{b. [\ion{N}{1}] vs. [\ion{O}{1}] $\lambda6300+$}
          }
\gridline{\fig{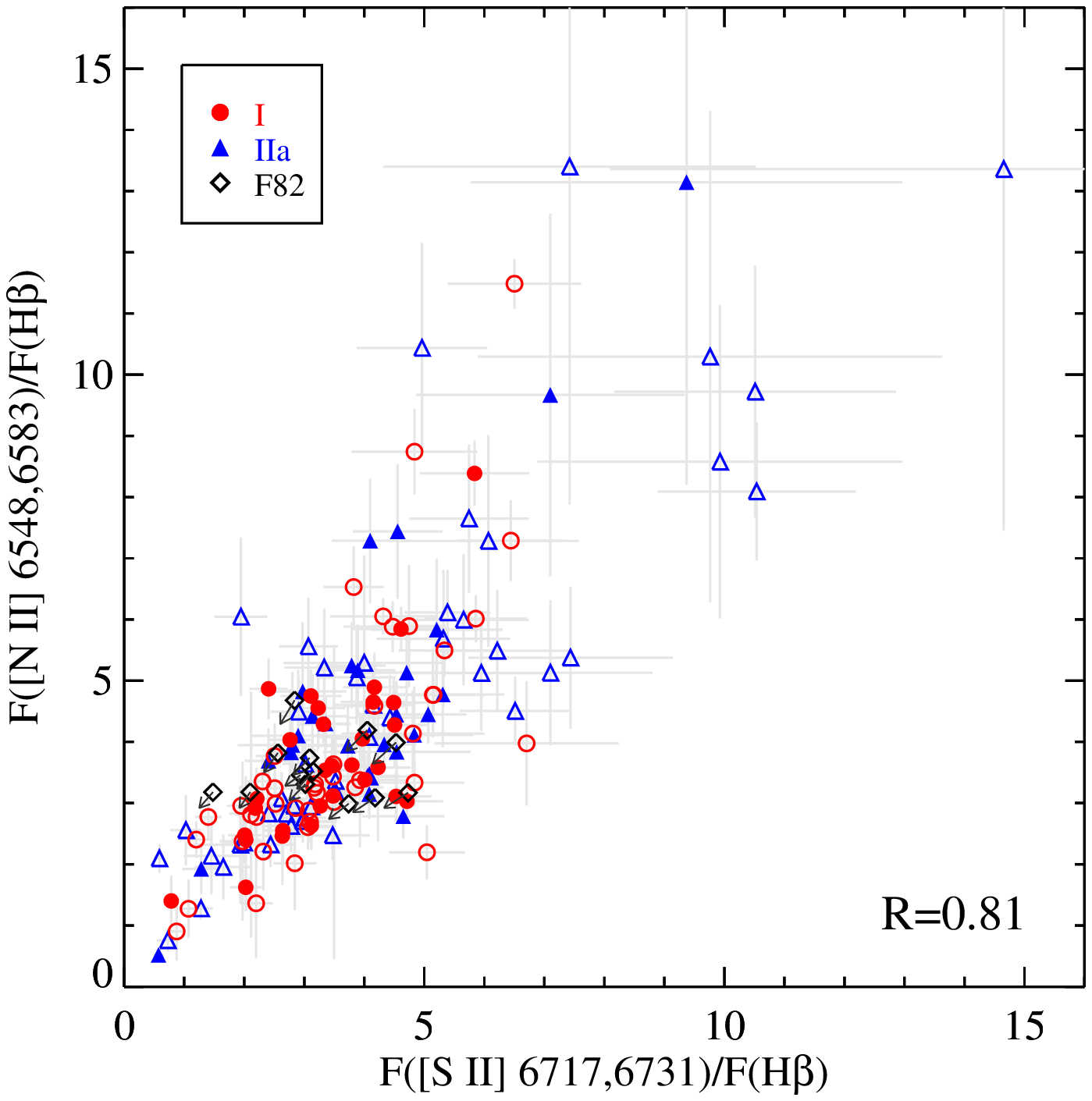}{0.5\textwidth}{c. [\ion{N}{2}] $\lambda6548+$ vs. [\ion{S}{2}] $\lambda6717+$}
          \fig{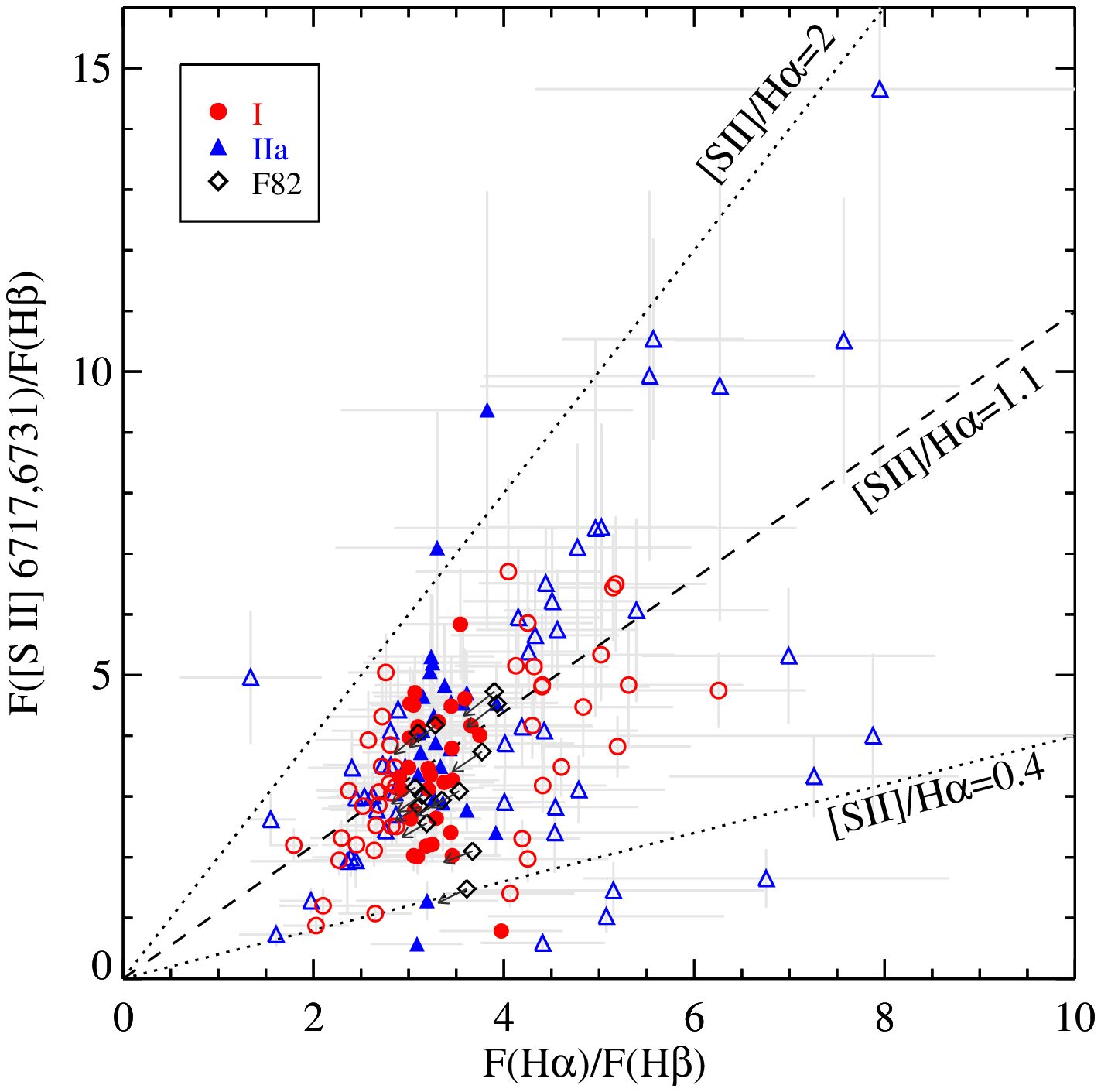}{0.5\textwidth}{d. [\ion{S}{2}] $\lambda6717+$ vs. H$\alpha$}
          }
    \caption{Correlation between line ratios of different elements: a. [\ion{Ne}{3}] vs. [\ion{O}{3}] $\lambda4959+$, b. [\ion{N}{1}] vs. [\ion{O}{1}] $\lambda6300+$, c. [\ion{N}{2}] $\lambda6548+$ vs. [\ion{S}{2}] $\lambda6717+$, and d. [\ion{S}{2}] $\lambda6717+$ vs. H$\alpha$. All line intensities are normalized to H$\beta$. Group I and IIa spectra are denoted with circles and triangles, respectively. Those with H$\alpha$/H$\beta<$2.9 or $>4.0$ are denoted with open symbol while the rest are marked with filled symbols (see text). For comparison, (measured) line ratios from \citetalias{fesen82} are overlaid with diamonds, and the effect of dereddening is marked with arrows. Hereafter, the symbol designation is applied in the same way for Figures \ref{fig:S2Ha}--\ref{fig:Te_N2}. For the correlations shown in panel a, b, and c, the correlation coefficients ($R$) are measured for Group I and IIa (those from \citetalias{fesen82} excluded). In panel d, a commonly used shock diagnostic, [\ion{S}{2}] $\lambda6717+$/H$\alpha$, is denoted. In most cases, [\ion{S}{2}] $\lambda6717+$/H$\alpha$ ranges between 0.4 and 2.0 (dotted lines), and the median ratio from our measurement is $\sim1.1$ (dashed line). }
    \label{fig:corr_elm}   
\end{figure}

\begin{figure}[!ht]
    \epsscale{1.15}
    \plottwo{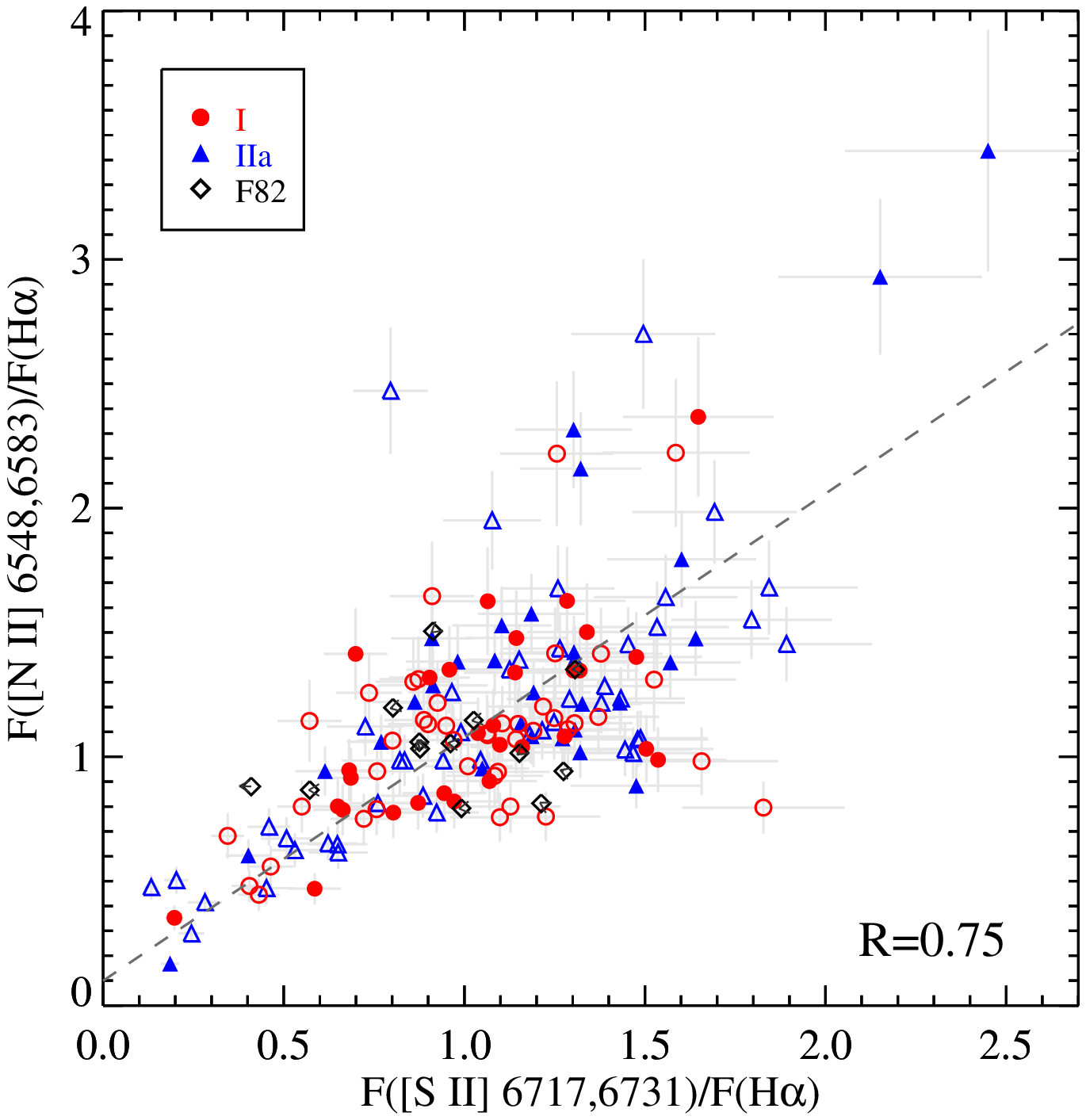}{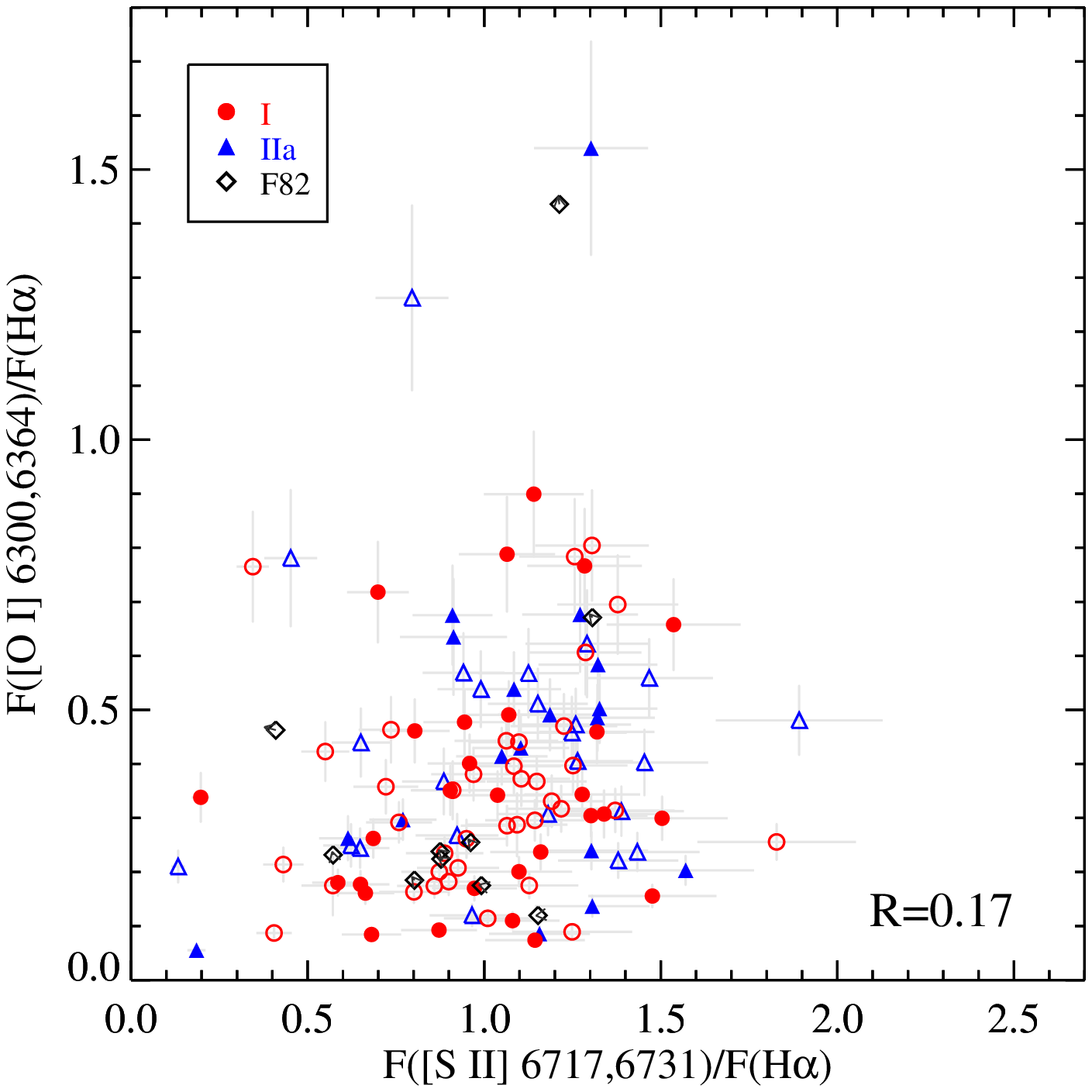}
    \caption{Shock diagnostic [\ion{S}{2}] $\lambda6717+$/H$\alpha$ ratios in comparison with [\ion{N}{2}] $\lambda6548+$/H$\alpha$ (left) and [\ion{O}{1}] $\lambda6300+$/H$\alpha$ (right). While [\ion{N}{2}] $\lambda6548+$/H$\alpha$ has a good correlation with [\ion{S}{2}] $\lambda6717+$/H$\alpha$ (correlation coefficient $R=0.75$), [\ion{O}{1}] $\lambda6300+$/H$\alpha$ shows no obvious evidence of correlation with [\ion{S}{2}] $\lambda6717+$/H$\alpha$ ($R=0.17$). For the correlation between [\ion{N}{2}] $\lambda6548+$/H$\alpha$ and [\ion{S}{2}] $\lambda6717+$/H$\alpha$, a linear fit is given with a dashed line ($y=a + bx$ where $a=0.10\pm0.04$ and $b=0.98\pm0.03$).     }
    \label{fig:S2Ha}
\end{figure}

Systematic correlations appear between line ratios of different elements, especially with the same ionization state. For examples, those with high ionization state such as [\ion{Ne}{3}] and [\ion{O}{3}] $\lambda4959+$ show a tight correlation (correlation coefficient\footnote{Note that the line ratios from \citetalias{fesen82} are not included for deriving correlation coefficients shown in Figures \ref{fig:corr_elm}--\ref{fig:corr_tran}.} of $R=0.90$) with each other (Figure \ref{fig:corr_elm}a). This is a natural consequence that lines from high ionization species tend to be strong where lines from other elements with high ionization are strong. Similarly, close correlations of [\ion{N}{1}] with [\ion{O}{1}] $\lambda6300+$ and [\ion{N}{2}] $\lambda6548+$ with [\ion{S}{2}] $\lambda6717+$ are also present (Figure \ref{fig:corr_elm}b, c). Where low ionization or neutral species emits strongly, lines from other low ionization species appear to be strong while lines from high ionization species become weaker.

The ratio of [\ion{S}{2}] $\lambda6717+$/H$\alpha$ is a well-known shock diagnostic \citep[e.g.,][]{math73}: a high [\ion{S}{2}] $\lambda6717+$/H$\alpha$ ratio (i.e., $\ga0.4$) indicating SNRs whereas a low ratio (often $\sim0.1$) for \ion{H}{2} regions. In Figure \ref{fig:corr_elm}d, the intensity of [\ion{S}{2}] $\lambda6717+$ is compared with the H$\alpha$ intensity. Except a few points, [\ion{S}{2}] $\lambda6717+$/H$\alpha$ ratios measured from the Group I and IIa spectra well exceed 0.4, corroborating the SNR origin. The observed ratios mostly range between 0.4 and 2.0 (between dotted lines in Figure \ref{fig:corr_elm}d), and the median is $\sim1.1$ (dashed line). Besides those with [\ion{S}{2}] $\lambda6717+$/H$\alpha\geq0.4$, there are 10 spectra (4 Group I and 6 Group IIa) showing significantly weak [\ion{S}{2}] $\lambda6717+$ emission relative to H$\alpha$ (i.e., $0.13\lesssim$[\ion{S}{2}] $\lambda6717+$/H$\alpha\lesssim0.34$). Half of them are along the interior filaments \citepalias[like position H of][]{fesen82}, four are located at the outskirts of the bright NE region NGC 6992, and one is near the bright SW region NGC 6960. These spectra show either strong [\ion{O}{1}] $\lambda6300+$ and/or [\ion{O}{2}] $\lambda3727$ line emission or high [\ion{O}{3}] $\lambda4959+$/H$\beta$ ratio, or both. Because of their locations as well as the spectral features, their emission is probably associated with the Cygnus Loop.  

We compare [\ion{N}{2}] $\lambda6548+$/H$\alpha$ and [\ion{O}{1}] $\lambda6300+$/H$\alpha$ ratios with respect to [\ion{S}{2}] $\lambda6717+$/H$\alpha$ in Figure \ref{fig:S2Ha}, which are previously known to correlate, especially for extragalactic SNRs \citep[e.g.,][]{smith93, gordon98, lee15, long18}. The [\ion{N}{2}] $\lambda6548+$/H$\alpha$ ratios of the Cygnus Loop show a fairly good correlation with [\ion{S}{2}] $\lambda6717+$/H$\alpha$ ratios ($R\simeq0.75$), verifying it as the secondary shock indicator. A linear fit to the correlation is performed, which gives [\ion{N}{2}] $\lambda6548+$/H$\alpha=(0.10\pm0.04) + (0.98\pm0.03) \times$[\ion{S}{2}] $\lambda6717+$/H$\alpha$ (dashed line in Figure \ref{fig:S2Ha}(left)). On the other hand, [\ion{O}{1}] $\lambda6300+$/H$\alpha$ shows no evidence for correlation with [\ion{S}{2}] $\lambda6717+$/H$\alpha$ ($R=0.17$, see Figure \ref{fig:S2Ha}(right)). This is somewhat surprising because [\ion{O}{1}] $\lambda6300+$ lines are considered to be a useful discriminant for shock-heated gas and this ratio shows as a good correlation with [\ion{S}{2}] $\lambda6717+$/H$\alpha$ as [\ion{N}{2}] $\lambda6548+$/H$\alpha$ for extragalactic SNRs \citep[e.g.,][see also \citealt{lee15}]{gordon98}. We attribute the lack of correlation partly to observational difficulties because the [\ion{O}{1}] emission from the night sky can contaminate the LAMOST spectra, especially those with low signal-to-noise ratios (see Tables \ref{tab:gr1}--\ref{tab:gr2a}). However, the correlation is not apparent in the samples of \citetalias{fesen82}, either. Also, the correlation appears weak for SNRs in some other galaxies \citep[see Figure 11 of][]{lee15}. Thus, the correlation between [\ion{O}{1}] $\lambda6300+$/H$\alpha$ and [\ion{S}{2}] $\lambda6717+$/H$\alpha$ may be limited to radiative SNRs with bright optical emission lines and needs a more careful investigation. 

\subsubsection{Line ratios of different transitions}

\begin{figure}[ht!]
    \gridline{\fig{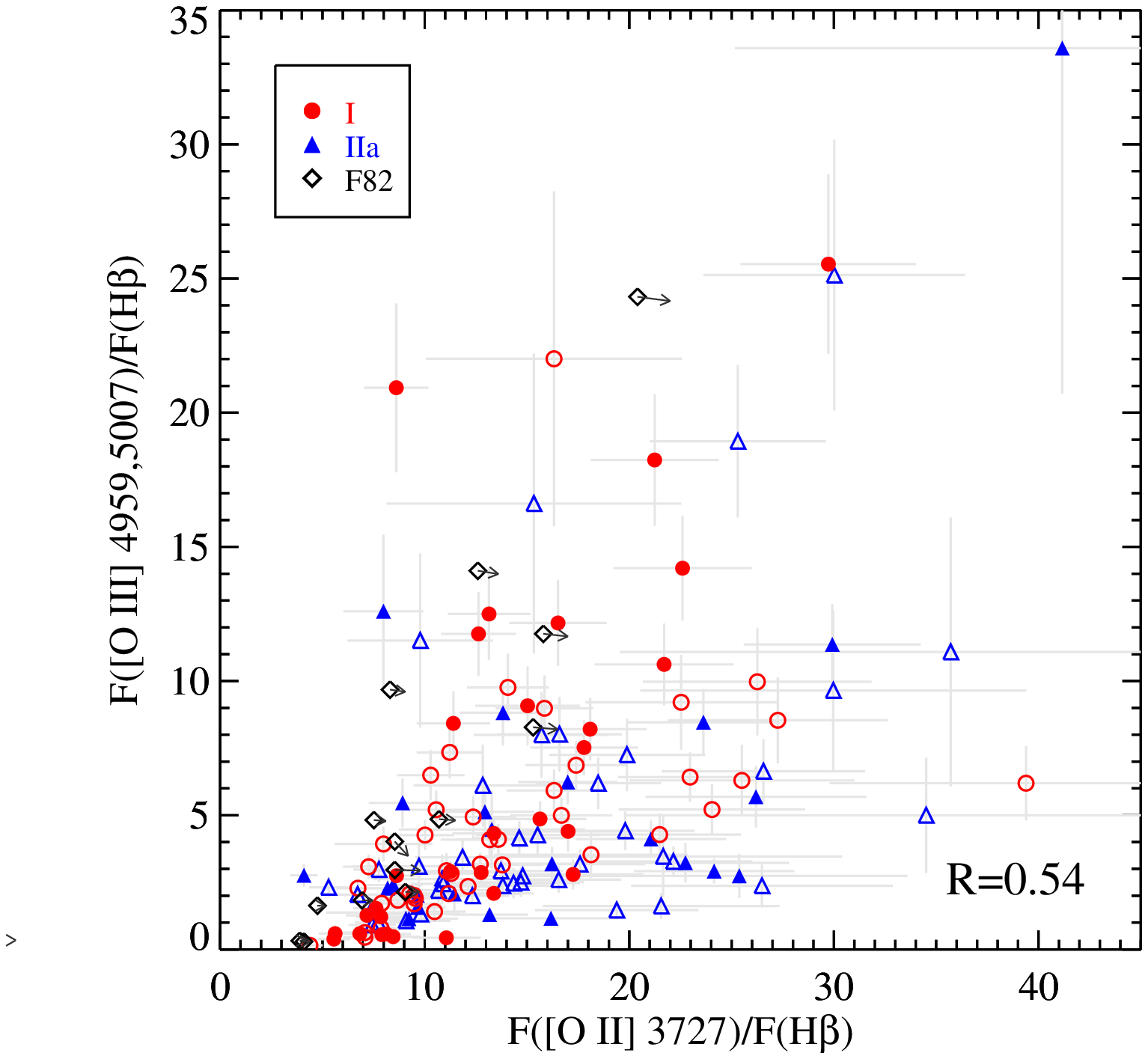}{0.5\textwidth}{a. [\ion{O}{3}] $\lambda4959+$ vs. [\ion{O}{2}] $\lambda3727$}
          \fig{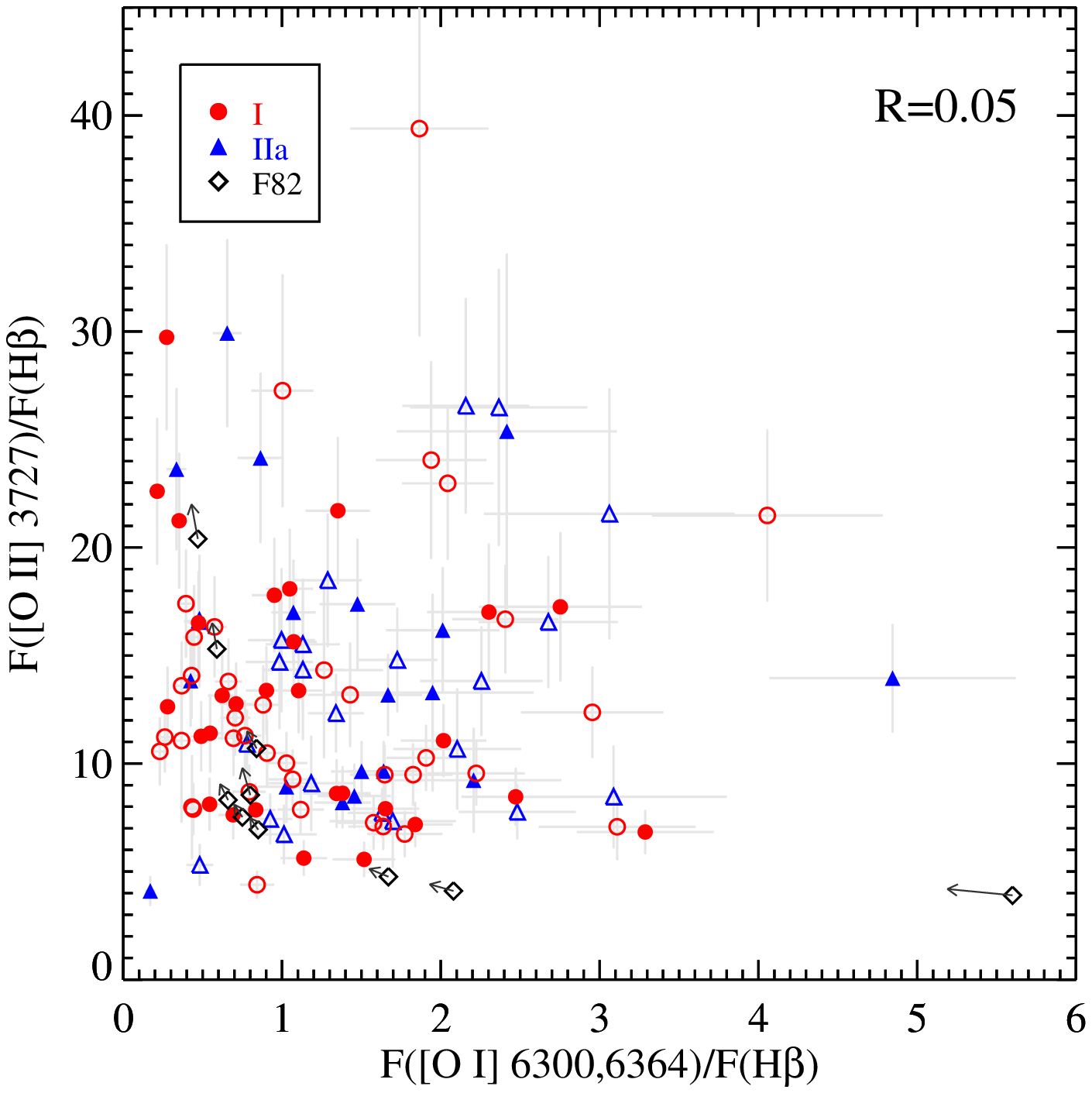}{0.5\textwidth}{b. [\ion{O}{2}] $\lambda3727$ vs. [\ion{O}{1}] $\lambda6300+$}
          }
    \gridline{\fig{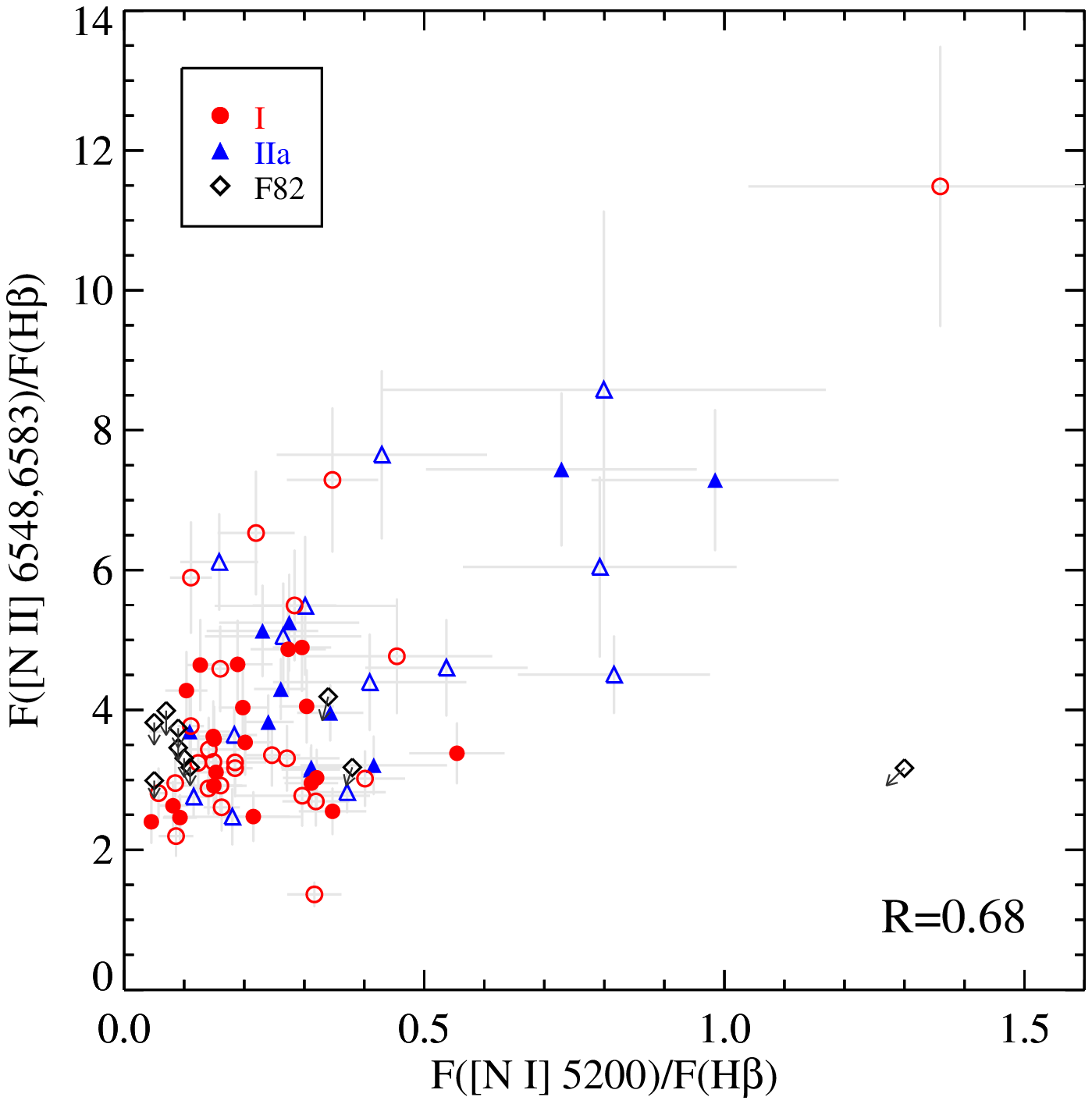}{0.5\textwidth}{c. [\ion{N}{2}] $\lambda6548+$ vs. [\ion{N}{1}]}
          \fig{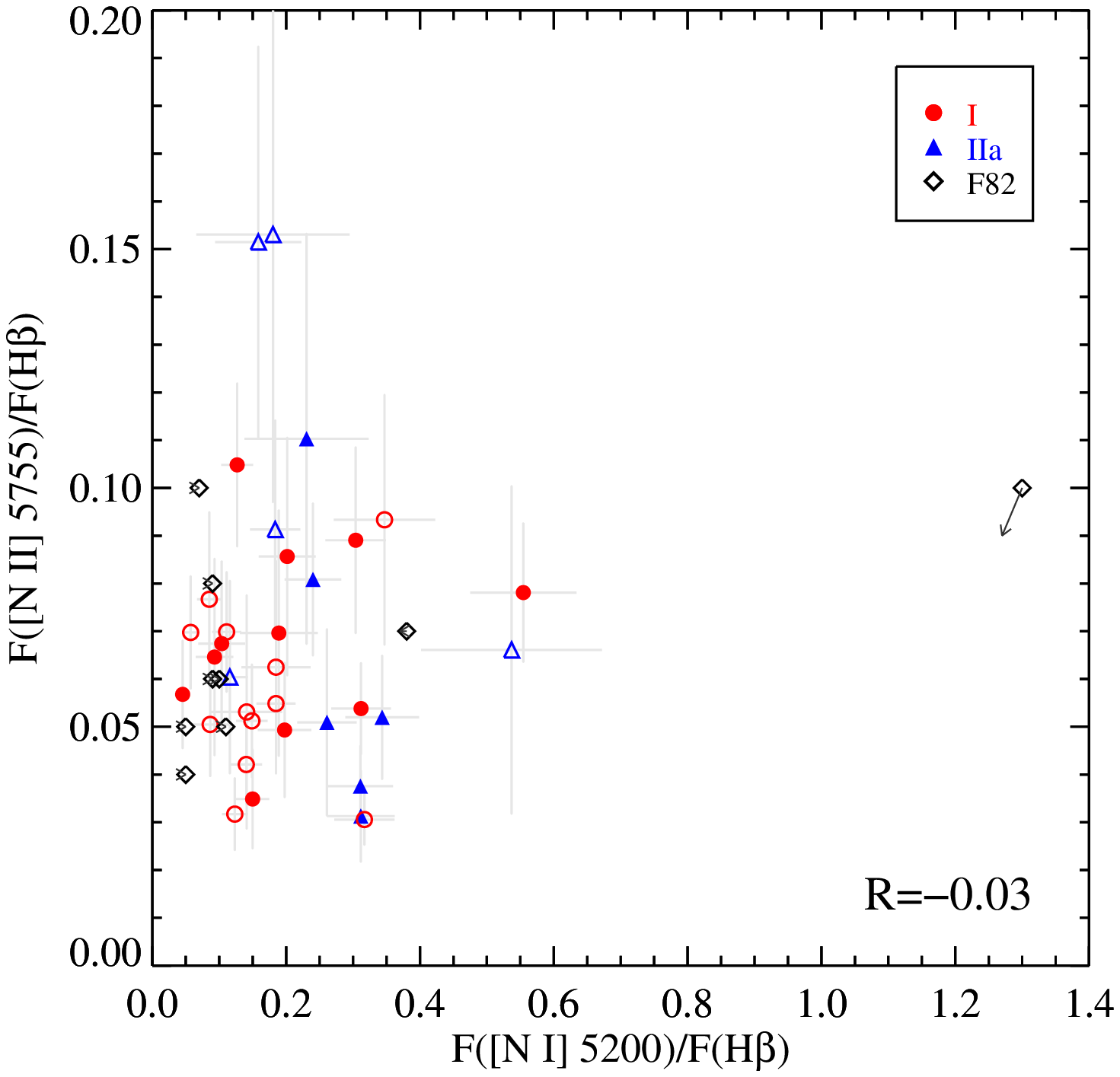}{0.5\textwidth}{d. [\ion{N}{2}] $\lambda5755$ vs. [\ion{N}{1}]}
          }
    \caption{Correlation between line ratios of the same elements (oxygen and nitrogen) with different transitions: a. [\ion{O}{3}] $\lambda4959+$ vs. [\ion{O}{2}] $\lambda3727$, b. [\ion{O}{2}] $\lambda3727$ vs. [\ion{O}{1}] $\lambda6300+$, c. [\ion{N}{2}] $\lambda6548+$ vs. [\ion{N}{1}], and d. [\ion{N}{2}] $\lambda$5755 vs. [\ion{N}{1}]. All line intensities are normalized to H$\beta$.}
    \label{fig:corr_tran}
\end{figure}

In Figure \ref{fig:corr_tran}, we examine line ratios of the same element such as oxygen or nitrogen with different transitions. The largest variation among all possible combination of line ratios is seen in the intensities of [\ion{O}{2}] $\lambda3727$ and [\ion{O}{3}] $\lambda4959+$ emission relative to H$\beta$, which range from $\sim0.15$ to $\sim40$. Such a large span clearly depicts the diversity of physical conditions within the Cygnus Loop \citep[e.g.,][]{fesen82, levenson98}. As previously reported in \citetalias{fesen82}, we also note that the ratio of [\ion{O}{2}] $\lambda3727$/H$\beta$ is no less than $\sim4$ in any case (see Figure \ref{fig:corr_tran}a, b) whereas both [\ion{O}{1}] $\lambda6300+$/H$\beta$ and [\ion{O}{3}] $\lambda4959+$/H$\beta$ can be as low as $\sim0.15$. This distinction of the [\ion{O}{2}] $\lambda3727$/H$\beta$ ratios (i.e., [\ion{O}{2}] $\lambda3727$/H$\beta\ga4$) has been seen in other Galactic SNRs as well as extragalactic SNRs \citep[e.g., see Figures 3 and 4 of][]{fesen85}, which can be used to separate SNRs from \ion{H}{2} regions. 

The intensities of [\ion{O}{2}] $\lambda3727$ and [\ion{O}{3}] $\lambda4959+$ relative to H$\beta$ appear to correlate moderately ($R=0.54$, Figure \ref{fig:corr_tran}a) whereas [\ion{O}{2}] $\lambda3727$ and [\ion{O}{1}] $\lambda6300+$ do not show an apparent correlation (Figure \ref{fig:corr_tran}b). However, as guided by the data from \citetalias{fesen82} (diamonds in Figure \ref{fig:corr_tran}b), the general trend would exist in a way that the [\ion{O}{2}] $\lambda3727$/H$\beta$ ratios tend to decrease as the [\ion{O}{1}] $\lambda6300+$/H$\beta$ ratios increase.

For nitrogen, the [\ion{N}{2}] $\lambda6548+$/H$\beta$ ratio seems to correlate with the [\ion{N}{1}]/H$\beta$ ($R=0.69$, Figure \ref{fig:corr_tran}c) while the [\ion{N}{2}] $\lambda5755$/H$\beta$ has no correlation with [\ion{N}{1}]/H$\beta$ ($R=-0.03$, Figure \ref{fig:corr_tran}d). Note that [\ion{N}{2}] $\lambda6548+$/$\lambda5755$ is sensitive to electron temperature (see Section \ref{subsec:Te_ne}). Then, the different trends seen in figure \ref{fig:corr_tran}c and \ref{fig:corr_tran}d imply the variation of temperature inside the remnant. However, we should be cautious to interpret the trends because those with large [\ion{N}{1}]/H$\beta$ ratios (i.e., [\ion{N}{1}]/H$\beta\ga0.7$) also have relatively large uncertainties, and those with the large [\ion{N}{1}]/H$\beta$ ratios (except one data point from \citetalias{fesen82}) do not appear in Figure \ref{fig:corr_tran}d due to non-detection of [\ion{N}{2}] $\lambda5755$ line. The presence of the correlation should be further examined with high signal-to-noise data. 

\clearpage

\subsection{Optical Properties: Electron Temperature and Density}\label{subsec:Te_ne}

\begin{figure}[ht!]
    \epsscale{1.1}
    \plotone{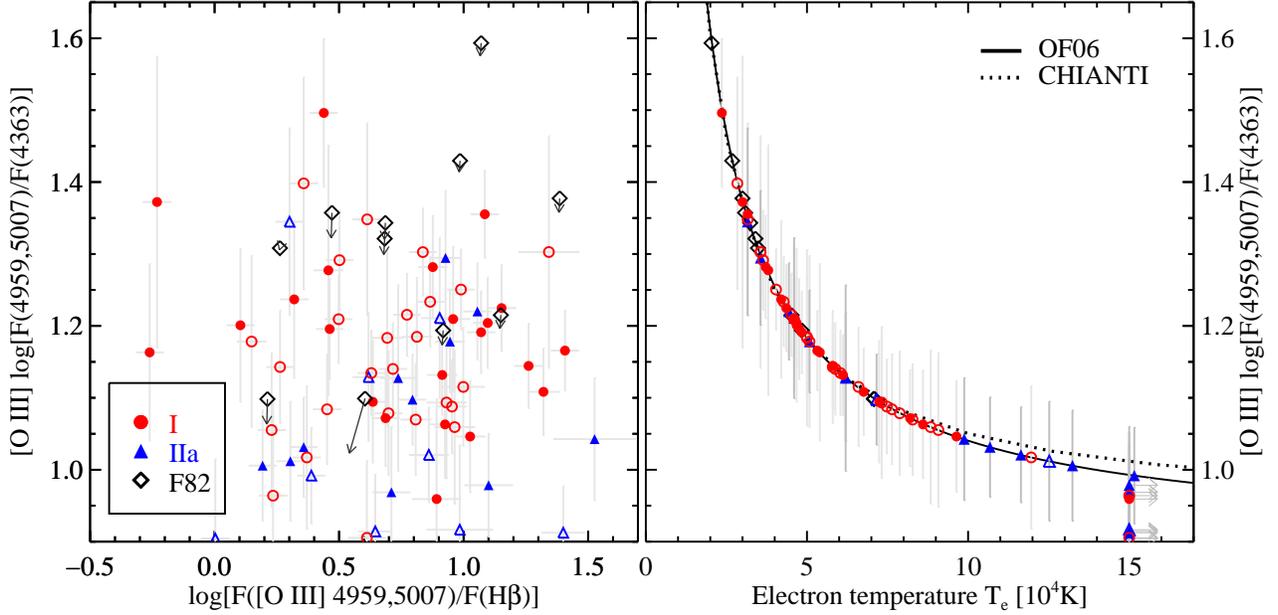}
    \caption{Left: comparison between [\ion{O}{3}] line temperatures and [\ion{O}{3}] $\lambda4959+$ line intensities relative to H$\beta$ for Group I and IIa spectra (circles and triangles, respectively). Data from \citetalias{fesen82} are also overlaid (diamonds). Right: electron temperatures derived from the [\ion{O}{3}] ratios using the exponential approximation of \citetalias{OF06} (solid line). CHIANTI model calculation is also overlaid (dotted line) for comparison. See Section \ref{subsec:Te_ne} for the details. }
    \label{fig:Te_O3}
\end{figure}

Electron temperatures of ionized plasma are commonly derived from a set of forbidden line emission emitted by metastable levels of positive ions, such as [\ion{O}{3}], [\ion{N}{2}], and [\ion{S}{2}]. A main representative is the ratio of [\ion{O}{3}] $\lambda$4959+/$\lambda4363$ \citep[e.g.,][OF06, hereafter]{OF06}. The [\ion{O}{3}] ratios in comparison with the [\ion{O}{3}] $\lambda4959+$ intensities relative to H$\beta$ are shown in Figure \ref{fig:Te_O3}(left). Adopting the exponential approximation of \citetalias{OF06} expressed as
\begin{equation} \label{eq:Te}
    \frac{j_{\lambda4959}+j_{\lambda5007}}{j_{\lambda4363}}=\frac{7.90 {\rm exp}(3.29\times10^4/T)}{1+4.5\times10^{-4}n_{\rm e}/T^{1/2}},
\end{equation}
[\ion{O}{3}] line temperatures for Group I and IIa are estimated in Figure \ref{fig:Te_O3}(right). In addition, we calculate theoretical ratios at a density of 100 cm$^{-3}$ using version 8 of the CHIANTI database \citep{dere97,delzanna15}, which are overlaid with a dotted line. CHIANTI consists of critically evaluated set of up-to-date atomic data, together with user-friendly programs written in Interactive Data Language (IDL) and Python to calculate the spectra from optically thin, collision-dominated astrophysical plasma\footnote{http://www.chiantidatabase.org}. Up to $\sim50$,000 K, the ratios from CHIANTI are consistent with those from the exponential approximation, but they start deviate at higher temperatures.

Overall, [\ion{O}{3}] temperatures from Group I and IIa spectra range between $\sim30,000$ K and 80,000 K, which are in good agreement with previous estimates \citep[e.g.,][]{miller74,fesen82}. However, a few cases with the [\ion{O}{3}] ratios less than $\sim10$ indicate that the temperature exceeds $\sim10^5$ K, which is above the equilibrium formation temperature. Such a high temperature has been reported previously \citep[e.g., $T_e\ga80$,000 K,][]{sankrit14}, which would occur in the narrow ionization zone just behind an X-ray producing (non-radiative) shock \citep[e.g.,][]{blair05}. The emission from these regions, however, could be too faint to be detected by LAMOST. Hence, there is a possibility that overestimation of the [\ion{O}{3}] $\lambda4363$ intensity leads the [\ion{O}{3}] ratios less than $\sim10$. Fitting its underlying baseline is sometimes uncertain due to the presence of absorption features nearby. In fact, Group IIa spectra which are more affected by stellar features tend to have lower [\ion{O}{3}] ratios than Group I, implying that the overestimation of [\ion{O}{3}] $\lambda4363$ is conceivable.

\begin{figure}[ht!]
    \epsscale{1.1}
    \plotone{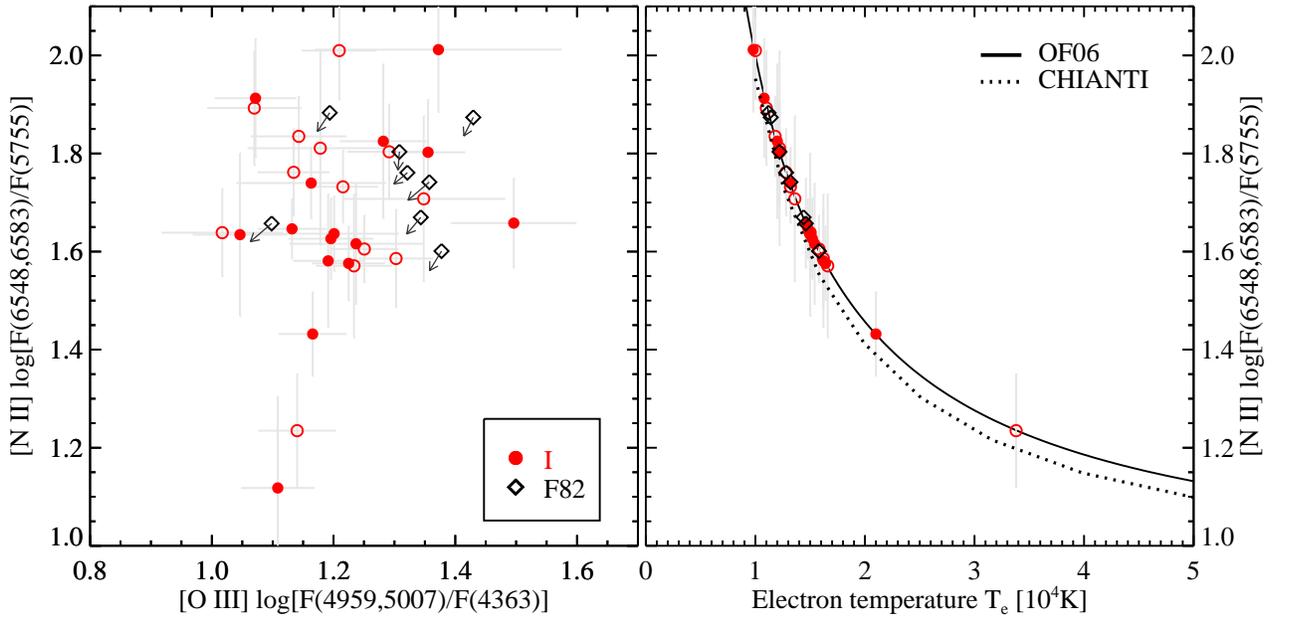}
    \caption{Left: comparison between [\ion{N}{2}] and [\ion{O}{3}] diagnostic of electron temperature. As [\ion{N}{2}] $\lambda$5755 line is mostly weak and contaminated in Group IIb spectra, only Group I is used here. Line ratios from \citetalias{fesen82} are also overlaid (diamonds). Right: electron temperatures derived from the [\ion{N}{2}] ratios using CHIANTI (dotted line). See Section \ref{subsec:Te_ne} for the details. }
    \label{fig:Te_N2}
\end{figure}

Another temperature diagnostic is the ratio of [\ion{N}{2}] $\lambda$6548+/$\lambda$5755. In Figure \ref{fig:Te_N2}, we compare the [\ion{N}{2}] and the [\ion{O}{3}] ratios and derive the [\ion{N}{2}] temperature in the same manner as Figure \ref{fig:Te_O3}. Again, adopting the exponential approximation of \citetalias{OF06}, the theoretical [\ion{N}{2}] ratio as a function of temperature ($T$) is given by
\begin{equation} \label{eq:TeNii}
    \frac{j_{\lambda6548}+j_{\lambda6583}}{j_{\lambda5755}}=\frac{8.23 {\rm exp}(2.50\times10^4/T)}{1+4.4\times10^{-3}n_{\rm e}/T^{1/2}},
\end{equation}
shown with a solid line in Figure \ref{fig:Te_N2}(right). Also, theoretical ratios from the CHIANTI database are overlaid with a dotted line. Note that the [\ion{N}{2}] $\lambda5755$ line is poorly detected in most Group IIa spectra due to its faintness, hence only Group I spectra are used to estimate [\ion{N}{2}] temperature. The resultant [\ion{N}{2}] temperatures are mostly between 10,000 and 15,000 K, which is substantially lower than those from [\ion{O}{3}] ratios (see Figure \ref{fig:Te_O3}). Also, there is no clear correlation between [\ion{N}{2}] temperatures and [\ion{O}{3}] temperatures. The higher temperature inferred from [\ion{O}{3}] with higher ionization state and no correlation between [\ion{N}{2}] and [\ion{O}{3}] are natural features for a region behind a radiating shock, where cooling and recombination to the lower ionization state occur in succession. This trend has been reported in literature \citep[e.g.,][]{miller74,fesen82} and is also found in other SNRs. For example, \citet{pauletti16} show the spatial variations in temperature maps of the SNR N49 in the Large Magellanic Cloud, which clearly demonstrate the higher temperatures for [\ion{O}{3}] line ratios compared to the [\ion{S}{2}], [\ion{O}{2}], and [\ion{N}{2}] temperatures and different spatial distribution of the temperatures through the SNR. 

\begin{figure}[ht!]
    \epsscale{1.1}
    \plotone{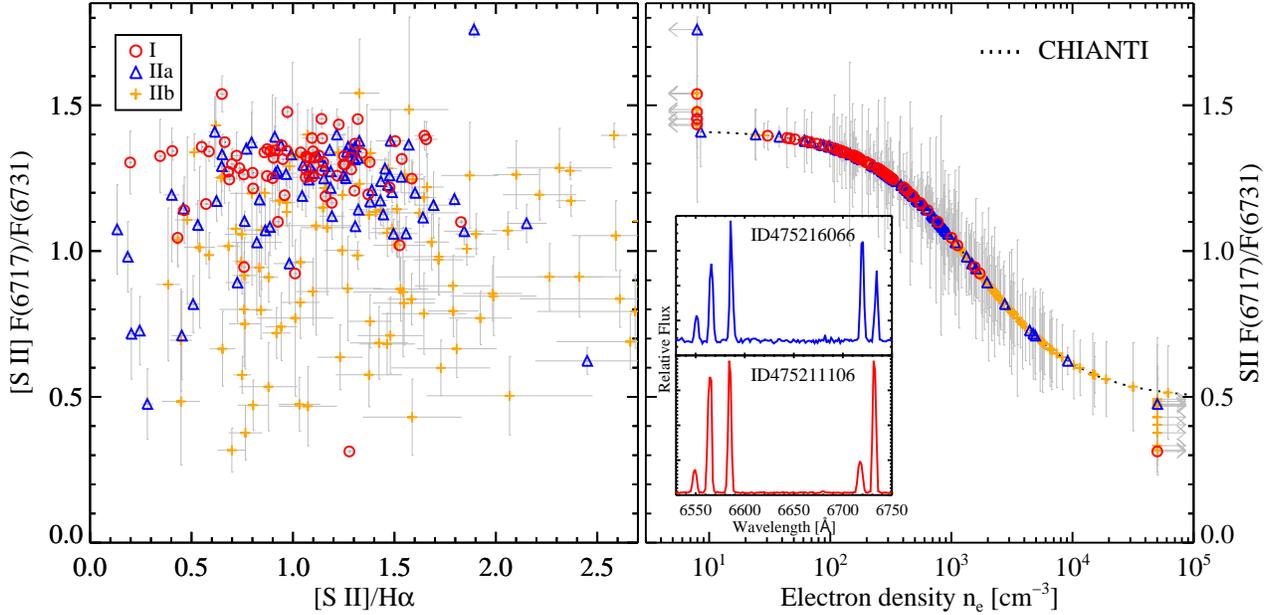}
    \caption{Left: [\ion{S}{2}] line ratio diagnostic of electron density in comparison with [\ion{S}{2}] $\lambda6717+$/H$\alpha$. Group I, IIa, and IIb spectra are denoted with circles, triangles, and crosses, respectively. Right: [\ion{S}{2}] electron density estimates by using CHIANTI. Two insets present the zoomed-in spectra showing two extreme cases (obs. ID 475211106 and 475216066) showing [\ion{S}{2}] ratio of $\sim0.31$ and 1.76, respectively. }
    \label{fig:ne}
\end{figure}

The line ratios of [\ion{S}{2}] $\lambda$6717/$\lambda$6731 \citep[e.g.,][]{OF06} and [\ion{O}{2}] $\lambda$3729/$\lambda$3726 \citep[e.g.,][]{pradhan06} are among common diagnostic tools for deriving the electron density ($n_{\rm e}$). As the latter pair is closely located in wavelength, they are not resolved in the LAMOST spectra ($R\sim1800$). 
In Figure \ref{fig:ne} (left), we compare the [\ion{S}{2}] $\lambda$6717/$\lambda$6731 ratios with the relative intensity of [\ion{S}{2}] $\lambda6717+$. Because the [\ion{S}{2}] doublet is clearly detected even in the Group IIb spectra in most cases, Group I, IIa, and IIb are used for $n_{\rm e}$ estimate. The [\ion{S}{2}] ratios mostly range between 1.0 and 1.5 while some outliers having extremely low or high values are present (see below). 

Using the CHIANTI calculations, we estimate the electron density from the [\ion{S}{2}] $\lambda$6717/$\lambda$6731 line ratios (Figure \ref{fig:ne}, right). [\ion{S}{2}] electron densities mostly range between $\lesssim20$ and $\approx500$ cm$^{-3}$, which are consistent with previous estimates \citep[e.g.,][]{miller74,fesen82}. Two significant outliers are the lowest ($\sim0.31$) and highest ($\sim1.76$) [\ion{S}{2}] ratios, which come from obs. ID 475211106 (Group I) and 475216066 (Group IIa) spectra as shown in the insets. The two spectra clearly show different trends: [\ion{S}{2}] $\lambda$6731 is much stronger than [\ion{S}{2}] $\lambda$6717 in obs. ID 475211106, and vice versa for obs. ID 475216066. The ratios outside the range given by the high ($\ga1.4$) and low ($\lesssim0.5$) density limits indicate measurement errors, which possibly result from the sky spectrum contaminated by the diffuse SNR emission or inapparent confusion with other emission sources (see more in Section \ref{subsec:disc}).

\section{Global Spectrum of the Cygnus Loop} \label{sec:global}

Using the Group I spectra, we have constructed a single integrated spectrum, which can represent a global spectrum of the Cygnus Loop. Because all the LAMOST spectra are only relatively flux calibrated (Section \ref{subsec:lamost}), absolute flux calibration is required to combine them. Generally, the absolute flux calibration needs spectra of standard stars under the same observing conditions. However, in the case of large spectroscopic surveys such as LAMOST, it is not straightforward to apply this strategy because it is impossible to obtain a sufficient number of spectra for standard stars every observing run. Thus, instead of precise absolute flux calibration, we have carried out crude flux calibration based on photometric magnitudes ($g$ and $i$ bands), which are already used for co-adding spectra with multi-exposures during the LAMOST pipeline \citep{du16}. Following Equations 4--6 in \citet{du16}, we have derived synthetic magnitudes and scale coefficients for each spectrum taking the Pan-STARRS1 $g$- and $i$-band transmission curves into account \citep{tonry12}. Then, the spectra are scaled using an average of the two scale coefficients and are accumulated into a single spectrum. After subtracting a continuum with a sixth-order polynomial fit, the final spectrum is obtained.  

\begin{figure}[ht!]
    \epsscale{1.0}
    \plotone{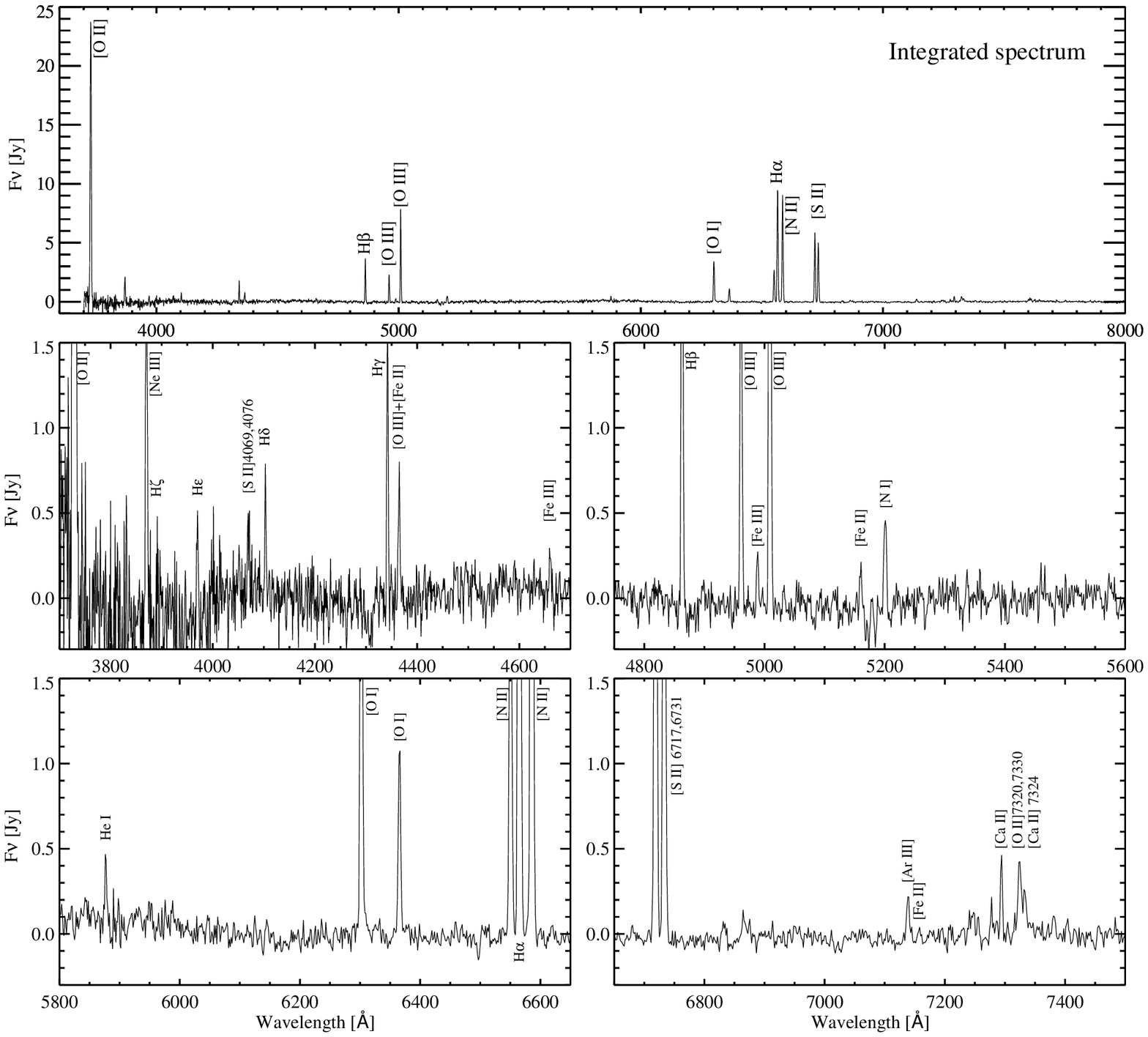}
    \caption{Global spectrum of the Cygnus Loop made of 75 Group I spectra (see section \ref{sec:global} for details). An entire spectrum (3600--8000 \AA) is shown in top panel while the other panels zoom in segments of the same spectrum to discern weaker lines. Noticeable lines are marked.}
    \label{fig:intspec}
\end{figure}

Figure \ref{fig:intspec} shows the global spectrum of the Cygnus Loop made by summing the 75 Group I spectra. The strongest emission in the spectrum is [\ion{O}{2}] $\lambda3727$, and several forbidden lines as well as the Balmer series clearly appear. Close-up views of the spectrum show the presence of weak lines (e.g., [\ion{Fe}{2}], [\ion{Fe}{3}], [\ion{Ar}{3}], [\ion{Ca}{2}], and \ion{He}{1}), too. In addition to the emission lines, contamination from stellar features (e.g., \ion{Mg}{1} triplet at 5167, 5172, and 5183 \AA) and residuals from imperfect sky subtraction (e.g., 6860--6960 \AA~due to telluric O$_2$) are also noticed. Intensities for detected emission lines are measured in the same way as described in Section \ref{subsec:line}, which are summarized in Table \ref{tab:line}. One of the main results seen in Table \ref{tab:line} is the moderate [\ion{O}{3}] $\lambda4959+$/H$\beta$ ratio of 2.98. Although the signature of incomplete shock (i.e., [\ion{O}{3}] $\lambda4959+$/H$\beta\ga$6) has been reported from a considerable number of positions in the remnant \citep[e.g.,][]{fesen82, raymond88}, our result indicates that a fully radiative shock is the most representative shock characteristic of the Cygnus Loop. This is not surprising because the global spectrum is inevitably predominated by bright emission regions, which usually arise from radiative shocks \citep[e.g.,][]{raymond88}. 

\begin{deluxetable}{ccr}\label{tab:line}

\tabletypesize{\footnotesize}


\tablecaption{Line Intensities of the Global Emission Spectrum of the Cygnus Loop (H$\beta$=100)}



\tablehead{\colhead{Ion} & \colhead{Wavelength} &  \colhead{Intensity} \\ 
 \colhead{ID}& \colhead{(\AA)}& \colhead{(relative to H$\beta$)}} 
\startdata
$[$\ion{O}{2}$]$ & 3727 & 1037  \\
$[$\ion{Ne}{3}$]$ & 3869 & 70\\
$[$\ion{O}{3}$]$ & 4363 & 21  \\
H$\beta$\tablenotemark{a} & 4864 & 100 \\
$[$\ion{O}{3}$]$ & 4959, 5007 & 298  \\
$[$\ion{N}{1}$]$ & 5200 & 20  \\
$[$\ion{N}{2}$]$ & 5755 & 6  \\
$[$\ion{O}{1}$]$ & 6300, 6364 & 175 \\
$[$\ion{N}{2}$]$ & 6548, 6584 & 443  \\
H$\alpha$ & 6564 & 379  \\
$[$\ion{S}{2}$]$ & 6717  & 223  \\
$[$\ion{S}{2}$]$ & 6730 & 192 \\
\hline 
\enddata
\tablenotetext{a}{Measured H$\beta$ flux is 1.34969 in counts.}



\end{deluxetable}

\subsection{Shock Parameters}

We investigate shock parameters to explain the measured line ratios of the global spectrum by using the shock code developed by \citet{raymond79} and \citet{cox85} with updated atomic parameters. Among the parameters necessary for the calculation of the forbidden lines of O and S, we updated the electron collision strengths to the recently calculated ones for \ion{O}{1} \citep{zatsarinny03}, \ion{O}{2} \citep{kisielius09}, \ion{O}{3} \citep{storey14}, and \ion{S}{2} \citep{tayal10}. The code assumes an 1D steady flow, using the Rankine-Hugoniot jump conditions to find the postshock gas parameters. Then it uses the fluid equations to compute the density, temperature, and velocity as the gas cools. The perpendicular component of the magnetic field is assumed to be frozen in, and it is compressed with the gas as it cools. Time-dependent ionization calculations including photoionization are used to compute the cooling rate and the emissivities of spectral lines.

Shock emission analysis of individual filaments in the Cygnus Loop has been carried out in several previous studies \citep[][and references therein]{miller74, raymond79, fesen82, hester83, raymond88, blair91, danforth01}. According to these studies, the optical spectra of bright filaments can be modeled by either complete or incomplete shocks with shock speeds in the range of 60--140~\kms\ and ambient densities 4--20~cm$^{-3}$. We have run shock models for shock speed $v_s = 60$--200~\kms\ and preshock density $n_0=10$~cm$^{-3}$. For the magnetic field strength $B_0$, we adopt 5~$\mu$G which is close to the median total magnetic field strength (6 $\mu$G) of the diffuse ($n\le 300$~cm$^{-3}$) interstellar cloud \citep{heiles05, crutcher10}. 
For the abundances of chemical elements, we use the solar abundances suggested by \citet{asplund09}, \citet{scott15a}, and \citet{scott15b}. The abundances of the elements that show strong lines in the global spectrum are [N/H]=7.83, [O/H]=8.69, [Ne/H]=7.93, 
and [S/H]=7.12 where [X/H] is log of number of X atoms per 
$10^{12}$ H atoms.
One complication in shock modeling is the preshock ionization levels of H and He that affect the postshock structure and therefore the emission line fluxes \citep{raymond79,shull1979,cox85,sutherland17}. 
We present a grid of models (Model F) where 
H is fully ionized and He is in ionization equilibrium with shock radiation.
The presence of neutral H would have an effect similar to that of lowering 
the shock velocity at full ionization \citep{cox85}. 
For comparison, we also present a grid of models (Model P) 
where H is partially ionized. In this model, the ionization fractions of 
H and He are determined by balancing the upstream ionizing flux with the incoming ion flux, which is 
a good approximation for slow shocks  \citep{shull1979,sutherland17}.
At \vsh$\geq 110$ \kms, H is fully ionized in model P, and the difference between the 
two models becomes negligible. 
Hence, we present Model F with \vsh=60--200 \kms, whereas Model P with 
\vsh=90--130 \kms~are used for comparison. Preshock ionization levels of the these cases are summarized in Table \ref{tab:ilevel}.
Finally, the models do not include emission from the photoionization precursor, which can be important for shocks faster than about 150 \kms\ \citep{dopita1996}. However, the precursor emission is faint and diffuse, so its contribution in a $2\farcs2$ fiber would be small.

\begin{deluxetable}{lcccccc|cccc}[hbt!]
\tablecaption{Input Parameter of Ionization Levels in Shock Models 
\label{tab:ilevel}}
\tablewidth{0pt}
\tablehead{
\colhead{} & \multicolumn{10}{c}{Shock Model} \\
\cline{2-11} \colhead{Parameter} & \colhead{F60} & \colhead{F80} & \colhead{F100} &\colhead{F120} &\colhead{F160} & \colhead{F200} & \colhead{P90} & \colhead{P100} & \colhead{P110}  &\colhead{P130}
}
\startdata
$v_{\rm s}$ (km s$^{-1}$) & 60 & 80 & 100 & 120 & 160 & 200 & 90 & 100 & 110 & 130 \\ 
preshock \ion{H}{1}  & 0.0 & 0.0 & 0.0 & 0.0 & 0.0 & 0.0 & 0.62 & 0.32 & 0.0 & 0.0 \\
preshock \ion{He}{1} & 0.84 & 0.28 & 0.07 & 0.03 & 0.02 & 0.0 & 0.95 & 0.66 & 0.0 & 0.0 \\
preshock \ion{He}{2} & 0.16 & 0.72 & 0.92  & 0.95  & 0.85 & 0.58 & 0.05 & 0.34 & 0.93 & 0.80 \\
\enddata
\tablecomments{H is fully ionized and He is in ionization equilibrium in Model F, whereas H is partially ionized in Model P. The ionization fractions of H and He in model P is from \citet{shull1979}.}
\end{deluxetable}

The measured line ratios are compared with the model calculations in Table \ref{tab:comp_ratio}. Considering that bright filaments in the Cygnus Loop are often assumed to have typical shock velocities around 100 \kms~in the literature,
most models in Table \ref{tab:comp_ratio} (i.e., \vsh$\ga$80 \kms) can reasonably reproduce the measurements within a factor of two or three. Models F120 and P110 show good agreement in the temperature-sensitive ratios (especially for [\ion{N}{2}] $\lambda6548+$/5755 ratios) consequently tracing shock velocity but predict slightly large [\ion{O}{3}] $\lambda4959+$/H$\beta$ and small [\ion{O}{2}] $\lambda3727+$/[\ion{O}{3}] $\lambda4959+$ ratios. In fact, all models except F60 and P90 produce lower [\ion{O}{2}]/[\ion{O}{3}] ratios than the observed one, and all but F60, P90, and F200 give higher [\ion{O}{3}]/H$\beta$ than the observed. This may indicate a mixture of low ($\lesssim100$ \kms) and high speed shocks with presence of partially ionized H.
In addition, the observed [\ion{O}{2}] $\lambda3727+$/[\ion{O}{3}] $\lambda4959+$ ratio higher than those shown in most of the shock models could result from depletion of carbon and silicon since [\ion{O}{2}] $\lambda3727+$/[\ion{O}{3}] $\lambda4959+$ is sensitive to these elemental abundances \citep{raymond79, fesen82}. 

Note that F120 or P110 models are not necessarily the best shock model to explain the global spectrum. Because we do not compare all measurable line ratios of between the data and the models, it could be unfair to choose the best shock model to describe the global properties of the Cygnus Loop just based on Table \ref{tab:comp_ratio}. However, the current results verify that the global spectrum can be characterized by fast (\vsh$\ga100$ \kms), radiative shocks and suggest the necessity of modifying the model parameters such as the elemental abundances. We will make detailed comparisons among different shock models and also discuss spatial variation of shock parameters in our forthcoming paper.

\begin{deluxetable}{cr@{$\pm$}lrrrrrr|rrrr}[!tbp]

\tabletypesize{\footnotesize}


\tablecaption{Line Ratios of the Global Spectrum with Shock Models\label{tab:comp_ratio}}



\tablehead{\colhead{Ratio} & \multicolumn2c{Observed} & \multicolumn{10}{c}{Shock Model} \\
\cline{4-13} \colhead{}& \multicolumn2c{value} &\colhead{F60} &\colhead{F80} & \colhead{F100} & \colhead{F120} & \colhead{F160} &\colhead{F200}  
 & \colhead{P90} & \colhead{P100}  & \colhead{P110} & \colhead{P130}}
\decimals
\startdata
$[$\ion{O}{3}$]$4959+/H$\beta$ & 2.98 & 0.39 & 0.13 & 5.19 & 5.09 & 4.91 & 3.86 & 2.93 & 0.28 & 3.78 & 4.96 & 5.24 \\
$[$\ion{O}{2}$]$3727+/$[$\ion{O}{3}$]$4959+ & 3.47 & 0.45 & 88.03 & 1.45 & 1.27 & 1.29 & 1.36 & 2.08 & 20.74 & 1.30 & 1.28 & 1.15 \\
$[$\ion{O}{3}$]$4959+/4363 & 14.1 & 2.3 & 17.94 & 17.59 & 17.80 & 17.99 & 15.89 & 16.65 & 17.32 & 17.42 & 18.04 & 16.74 \\ 
$[$\ion{N}{2}$]$6548+/5755 & 71.4 & 6.0 & 34.18 & 40.83 & 59.62 & 77.71 & 100.88 & 110.11 & 35.43 & 48.55 & 69.23 & 79.39 \\
$[$\ion{N}{2}$]$6548+/$[$\ion{O}{2}$]$3727 & 0.43 & 0.06 & 0.16 & 0.20 & 0.29 & 0.38 & 0.44 & 0.48 & 0.16 & 0.24 & 0.34 & 0.39\\
H$\alpha$/$[$\ion{N}{2}$]$6548+ & 0.86 & 0.12 & 1.81 & 2.20 & 1.62 & 1.23 & 1.29 & 0.99 & 3.21 & 2.54 & 1.41 & 1.27 \\
H$\alpha$/$[$\ion{S}{2}$]$6717+ & 0.92 & 0.12 & 2.13 & 2.08 & 1.13 & 0.84 & 1.05 & 0.84 & 3.75 & 2.05 & 0.95 & 0.91 \\
$[$\ion{S}{2}$]$ 6717/6731 & 1.16 & 0.18 & 1.31 & 1.25 & 1.24 & 1.22 & 1.09 & 1.07 & 1.23 & 1.22 & 1.24 & 1.19 \\
\enddata


\end{deluxetable}

\subsection{Discussion}\label{subsec:disc}

One of the main results that the LAMOST data show is that the line intensities inside the remnant vary more significantly than was previously thought, perhaps because earlier studies selected bright filaments. The uncertainties in the LAMOST data that cannot be explicitly estimated would account for some of the variation (see below). However, the large variation in the line ratios can still have an important impact on understanding evolutionary stages of SNRs as well as characteristics of extragalactic SNRs, particularly because small variation in line strength within a single SNR is often a fundamental assumption for these studies \citep[e.g.,][]{daltabuit76, fesen85}. The most commonly used ratios for that purpose are H$\alpha$/[\ion{N}{2}] $\lambda6548+$, H$\alpha$/[\ion{S}{2}] $\lambda6717+$, and [\ion{S}{2}] $\lambda$6717/$\lambda$6731 \citep[e.g.,][]{blair85,fesen85,lee15,winkler17}; The former two ratios probe the N/H and (to some degree) S/H abundances, consequently representing local metallicity, and the [\ion{S}{2}] doublet ratio is a well-known diagnostic of electron density. As the total number of spectroscopic pointings inside the Cygnus Loop increases more than an order of magnitude compared to previous studies, it would be meaningful to provide new ranges of these line ratios and to revisit their trends. 

\begin{figure}[!tbp]
    \epsscale{1.1}
    \plotone{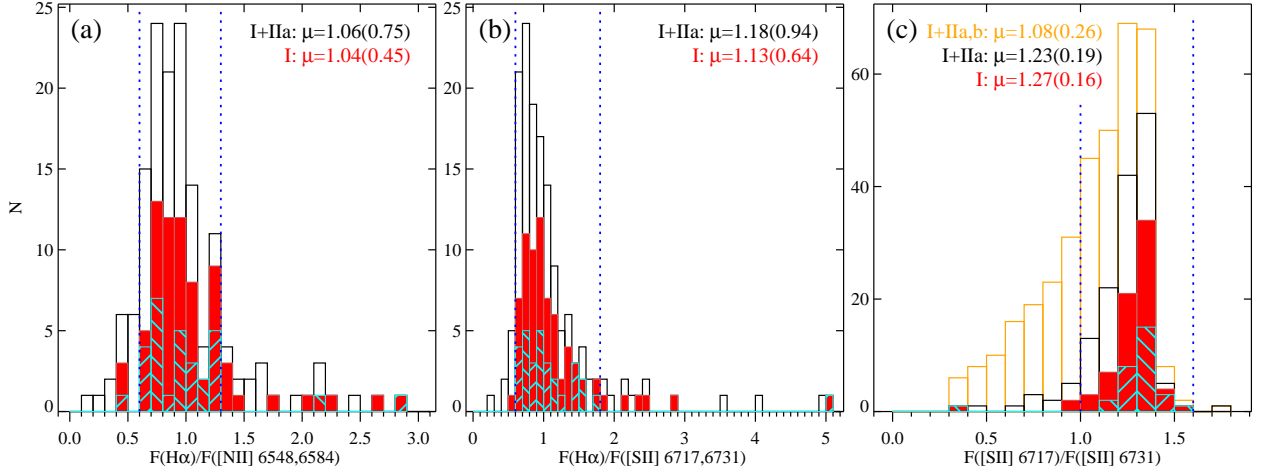}
    \caption{Line intensity variations inside the Cygnus Loop: (a) H$\alpha$/[\ion{N}{2}] $\lambda6548+$, (b) H$\alpha$/[\ion{S}{2}] $\lambda6717+$, and (c) [\ion{S}{2}] $\lambda$6717/$\lambda$6731. Ratio distributions of Group I and IIa (black lines) and Group I only (red bars) are presented in all panels, and Group IIb (yellow lines) are also included in the panel (c). Those in Group IIa having extremely large ratios are not shown but included in all analysis. A subset of Group I that has $2.9\leq$H$\alpha$/H$\beta\leq4.0$ is differentially marked (cyan shade). Two dotted lines indicate the minimum and maximum ratios that have been reported in the literature (see text). The mean ($\mu$) of each ratio with standard deviation (in parenthesis) is noted in each panel. }
    \label{fig:hist_ratio}
\end{figure}

Figure \ref{fig:hist_ratio} shows histogram distributions of H$\alpha$/[\ion{N}{2}] $\lambda6548+$, H$\alpha$/[\ion{S}{2}] $\lambda6717+$, and [\ion{S}{2}] $\lambda$6717/$\lambda$6731. Group I and IIa spectra are included for all cases, and Group IIb are also used for the [\ion{S}{2}] doublet the same as Figure \ref{fig:ne} shows. The distributions of Group I spectra (red bars in Figure \ref{fig:hist_ratio}) show that the ranges of the H$\alpha$/[\ion{N}{2}] $\lambda6548+$, H$\alpha$/[\ion{S}{2}] $\lambda6717+$, and [\ion{S}{2}] $\lambda$6717/$\lambda$6731 ratios are 0.42--2.84, 0.55--5.07, and 0.31--1.54, respectively. When Group IIa are included, these ranges increase by a factor of 2--3 whereas the case of [\ion{S}{2}] doublet does not show much change even if Group IIb are included. This suggests that the presence of Group IIa outliers that significantly increase the ratio range are likely due to some errors resulting from imperfect subtraction of the H$\alpha$ absorption feature. In addition, we also note that the ratio range of those with $2.9\leq$H$\alpha$/H$\beta\leq4.0$ in Group I (i.e. excluding those with large uncertainties in the Balmer line ratio, Section \ref{subsec:line}) is as wide as that of all Group I. In fact, it is a natural consequence that uncertainties related to H$\alpha$/H$\beta$ such as mismatch between blue and red spectra or any calibration errors depending on wavelength cannot affect these ratios significantly as H$\alpha$, [\ion{N}{2}] $\lambda6548+$, and [\ion{S}{2}] $\lambda6717+$ lines are located very closely.  

\citet{fesen85} collected previous observational results about these ratios in several Galactic SNRs (see their Table 5). The minimum and maximum values of each ratio combining all previous studies of the Cygnus Loop in their table are 0.66--1.25, 0.61--1.76, and 1.00--1.51 for H$\alpha$/[\ion{N}{2}] $\lambda6548+$, H$\alpha$/[\ion{S}{2}] $\lambda6717+$, and [\ion{S}{2}] doublet, respectively (marked with dotted lines in Figure \ref{fig:hist_ratio}). Note that the largest number of observations included in \citet{fesen85} is 18 \citep{parker64} while the number of Group I and IIa spectra are 75 and 79, respectively. We examine the Group I spectra that give significantly large H$\alpha$/[\ion{N}{2}] $\lambda6548+$ ($\ga1.5$) and H$\alpha$/[\ion{S}{2}] $\lambda6717+$ ($\ga2.0$) ratios. There are six and five Group I spectra with such large H$\alpha$/[\ion{N}{2}] $\lambda6548+$ and H$\alpha$/[\ion{S}{2}] $\lambda6717+$, respectively, and four of them are in common. All these outliers except one (obs. ID 470503149) show strong [\ion{O}{3}] $\lambda$4959+ emission relative to H$\beta$, and more than half have [\ion{O}{3}] $\lambda4959+$/H$\beta\ga6$ implying their association with incomplete shocks. It is clear that the line ratios resulting from the LAMOST data are more diverse than those in the literature although a part of this diversity is due to the errors in the LAMOST ratios. For [\ion{S}{2}] $\lambda6717/\lambda6731$, most of Group I (and IIa) spectra well agree with the previous range except the one (obs. ID 475211106) as noticed in Figure \ref{fig:ne}. This is reasonable because its variation is tightly constrained by electron density. As mentioned in Section \ref{subsec:Te_ne}, however, the spectrum of obs. ID 475211106 is problematic since its [\ion{S}{2}] doublet ratio is smaller than the high density limit (i.e. lower than 0.5). It is difficult to explain such a low ratio by any common errors including calibration, data reduction, and background confusion, because the emission lines including [\ion{S}{2}] doublet in that spectrum are clearly detected with high S/N and their line profiles are also well-shaped (i.e., no possible residuals from sky subtraction). Further observations with high spatial precision and high spectral resolution are needed to clarify the origin of this abnormal ratio.

Mean values $\mu$ (standard deviation) of the ratios are 1.04 (0.45), 1.13 (0.64), and 1.27 (0.16) for H$\alpha$/[\ion{N}{2}] $\lambda6548+$, H$\alpha$/[\ion{S}{2}] $\lambda6717+$, and [\ion{S}{2}] doublet, respectively when Group I are only considered. These values are changed when Group IIa (and IIb) are included, but the change is not significant. Corresponding mean values listed in \citet{fesen85} range 0.88--0.99, 1.00--1.08, and 1.19--1.40, respectively, which well agree with the newly measured $\mu$ despite the diversity of the ratio ranges that Group I (and IIa) show. In other words, although the standard deviations of the line ratios are larger than the previous measurements, their mean values are overall consistent. This result implies that as the number of observations (i.e., area that spectroscopy covers) increases, the range of the line ratios might widen, but their mean values can remain the same. This supports the validity of these line ratios as a probe of the evolutionary state or as a tracer of the elemental abundance of the ambient medium.  

Another aspect that the LAMOST data, particularly those from the faint filaments, show is the possible contribution of background emission including the precursor emission, the Galactic H$\alpha$ emission, and the Geocoronal H$\alpha$. By targeting bright filaments in the Cygnus Loop, previous studies \citep[e.g.,][and references therein]{fesen82, fesen85} can consequently minimize (and subtract off) the background contribution in their sample spectra. The slightly lower H$\alpha$/[\ion{N}{2}] $\lambda6548+$ and H$\alpha$/[\ion{S}{2}] $\lambda6717+$ ratios reported in \citet{fesen85} than those derived from the LAMOST data could be explained by this. On the contrary, the spectra of poorly resolved (e.g., the Magellanic Clouds) or unresolved (other distant galaxies) SNRs can be affected by these background sources more significantly. In particular, the precursor emission is very diffuse, so its contribution to a global spectrum of an extragalactic SNR would not be negligible compared to bright filaments of the Cygnus Loop or any other bright filaments of Galactic SNRs studied earlier. We will examine the effect of the precursor using shock models in the forthcoming paper. 

It is worthwhile to point out that the corresponding ratios of the global spectrum, shortly global ratios, (0.86, 0.92, and 1.16, see Table \ref{tab:comp_ratio}) are systematically lower than the $\mu$ values. In particular, since a single spectrum obtained from a spatially unresolved SNR in an external galaxy would be analogous to the global spectrum, it is critical to understand features of the global spectrum and how to interpret them correctly. The global spectrum can be considered to be the brightness-weighted summation while $\mu$ is a result of unweighted summation. When outliers have extremely large or small ratios, this can affect $\mu$ regardless of their brightness but less so for the global spectrum if the outliers are rather faint. On the other hand, the global spectrum is always dominated by spectra with high surface brightness. In the case of Group I (and IIa), there are particularly large H$\alpha$/[\ion{N}{2}] $\lambda6548+$ and H$\alpha$/[\ion{S}{2}] $\lambda6717+$ ratios so that the $\mu$ values become relatively larger than those of the global spectrum. Interestingly, the [\ion{S}{2}] doublet ratio of the global spectrum is smaller than $\mu$ of Group I, indicating that the bright Group I spectra tend to have small [\ion{S}{2}] ratios tracing high density regions. This is consistent with the aforementioned description of the global spectrum, which represents the bright spectra usually emitted by dense material swept-up by radiative shocks. 

To avoid the effect of the outliers, a median value of each ratio is also examined: 0.92. 0.94, and 1.31 for H$\alpha$/[\ion{N}{2}] $\lambda6548+$, H$\alpha$/[\ion{S}{2}] $\lambda6717+$, and [\ion{S}{2}] doublet, respectively in the case of Group I spectra. The smaller values for H$\alpha$/[\ion{N}{2}] $\lambda6548+$, H$\alpha$/[\ion{S}{2}] $\lambda6717+$ and larger one for [\ion{S}{2}] doublet compared to $\mu$ are a natural consequence of the outliers. These median values can be interpreted as the most common ratios from the sample spectra. Comparing the median ratios with the global ratios, the H$\alpha$/[\ion{N}{2}] $\lambda6548+$ and H$\alpha$/[\ion{S}{2}] $\lambda6717+$ ratios show better consistency than $\mu$, but the global [\ion{S}{2}] ratio is again smaller than the median. The better agreement seen in the former two line ratios could support their validity, meaning that these ratios from the global spectrum can probe the overall abundance of the SNR. This can be reasonable if emitting material inside an SNR is mostly ambient ISM with uniform abundance. However, in the case of young core-collapse SNRs where newly-formed ejecta can significantly contribute, a global spectrum might give a misleading value for the abundances of the SNR. In the case of [\ion{S}{2}] doublet, it again shows that the ratio of the global spectrum is smaller than the median value. Hence, this further suggests that the electron density measured from the global spectrum is likely to be biased toward denser regions.

\section{Summary} \label{sec:sum}

We have examined the prototypical middle-aged SNR, Cygnus Loop using unbiased spectroscopic data obtained with LAMOST. Both its large field-of-view ($\sim20$ deg$^2$) nearly as large as the size of the Loop and the multi-object spectrographs that can obtain 4000 spectra simultaneously provide a unique opportunity to spectroscopically study the entire SNR en masse. In the field of the Cygnus Loop, 2747 spectra are found in the LAMOST DR5, and 368 spectra are confirmed to exhibit emission lines originating from the SNR. In this paper, we describe the basic information on the LAMOST data and the classification of the spectra, and examine correlation of line ratios and the global spectrum of the SNR. The primary results are as follows.

1. Based on the presence of emission lines associated with the SNR and the contamination from background/foreground stars, 75, 79, and 214 spectra are classified into Group I, IIa, and IIb, which represent SNR-dominated emission, clear SNR emission with stellar features, and relatively weak SNR emission with dominant stellar features, respectively. Besides, 176 spectra exhibit emission lines of which origin is inconclusive (categorized into Group III). As the spatial distribution of this Group is mostly near the bright filaments, it is likely that the Group III spectra are also associated with the Cygnus Loop.

2. Combining the 75 Group I and 79 Group IIa spectra, the 154 spectra are further examined in detail. Various emission lines are identified, and relative intensities of the key lines are measured. The relative strengths of line emission show the spatial variation; In particular, wide ranges of [\ion{O}{3}] $\lambda$4959+/H$\beta$ and other line ratios such as [\ion{O}{2}] $\lambda$3727/H$\beta$, [\ion{N}{2}] $\lambda$6548+/H$\beta$ indicate the diversity of the physical parameters coexisting inside the single SNR. 

3. Line ratios of different elements with the same ionization state generally show systematic correlations. The [\ion{S}{2}] $\lambda6717+$/H$\alpha$ ratio, a well-known shock diagnostic, appear to correlate well with [\ion{N}{2}] $\lambda6548+$/H$\alpha$, whereas [\ion{O}{1}] $\lambda6300+$/H$\alpha$ show no clear evidence of correlation. This implies that [\ion{N}{2}] $\lambda6548+$/H$\alpha$ is more reliable secondary shock tracer than [\ion{O}{1}] $\lambda6300+$/H$\alpha$. 

4. Electron temperatures estimated with the [\ion{O}{3}] ratio mostly range between $\sim$3--8$\times10^4$ K while those with the [\ion{N}{2}] ratio range between $\sim1$--1.5$\times10^4$ K. The difference between the two estimates is a natural feature for a region behind a radiative shock, where cooling and recombination to the lower ionization state occur in succession. The electron density of the Cygnus Loop is mostly between 20 and 500 cm$^{-3}$ while some outliers indicate observational uncertainties.

5. The global spectrum of the Cygnus Loop demonstrates characteristics of a fully radiative shock albeit the presence of incomplete shocks inside the remnant. Comparison between the line ratios of the global spectrum and shock models verifies that fast (\vsh=100--140 \kms), radiative shocks can explain the observed ratios reasonably well but also suggests local variations of the shock parameters as well as the possible depletion of carbon and silicon. 

6. Group I and IIa spectra show wider ranges of the line ratios (H$\alpha$/[\ion{N}{2}] $\lambda6548+$, H$\alpha$/[\ion{S}{2}] $\lambda6717+$, and [\ion{S}{2}] $\lambda6717/\lambda6731$) than those previously reported. This implies that local variation in physical properties inside a single SNR can be more significant than commonly assumed, though uncertainties in the fluxes also contribute. In addition, the median values of the former two ratios are consistent with the corresponding ratios derived from the global spectrum while the median of the [\ion{S}{2}] doublet is larger than that from the global ratio. These results suggest that an optical spectrum of an unresolved, extragalactic SNR can probe its overall elemental abundance reasonably well, while its density diagnostics tend to overestimate its density. 

In our forthcoming papers, we will make detailed comparisons among different shock models, examine spatial variation of shock properties inside the Cygnus Loop, and perform analysis of kinematics. A combination of low/medium-resolution multi-object spectrograph with a large field-of-view and multi-wavelength imaging surveys including the SNR and its neighbouring regions will complete our understanding of the Cygnus Loop in a large scale and will benefit the interpretation of distant SNRs.  

\acknowledgements
JYS and GZ were supported by NSFC Grant Nos. 11988101, 11650110436, and 11890694, and JYS thanks the Chinese Academy of Sciences (CAS) for support through LAMOST fellowship. BCK acknowledges support from the Basic Science Research Program through the National Research Foundation of Korea (NRF) funded by the Ministry of Science, ICT and future Planning (2017R1A2A2A05001337). Funding for LAMOST (http://www.lamost.org) has been provided by the National Development and Reform Commission. LAMOST is operated and managed by the National Astronomical Observatories, Chinese Academy of Sciences. CHIANTI is a collaborative project involving George Mason University, the University of Michigan (USA), University of Cambridge (UK) and NASA Goddard Space Flight Center (USA).

\appendix


\figsetstart
\figsetnum{A. 1}
\figsettitle{Full LAMOST data of 75 Group I spectra}

\figsetgrpstart
\figsetgrpnum{1.1}
\epsscale{1.1}
\figsetplot{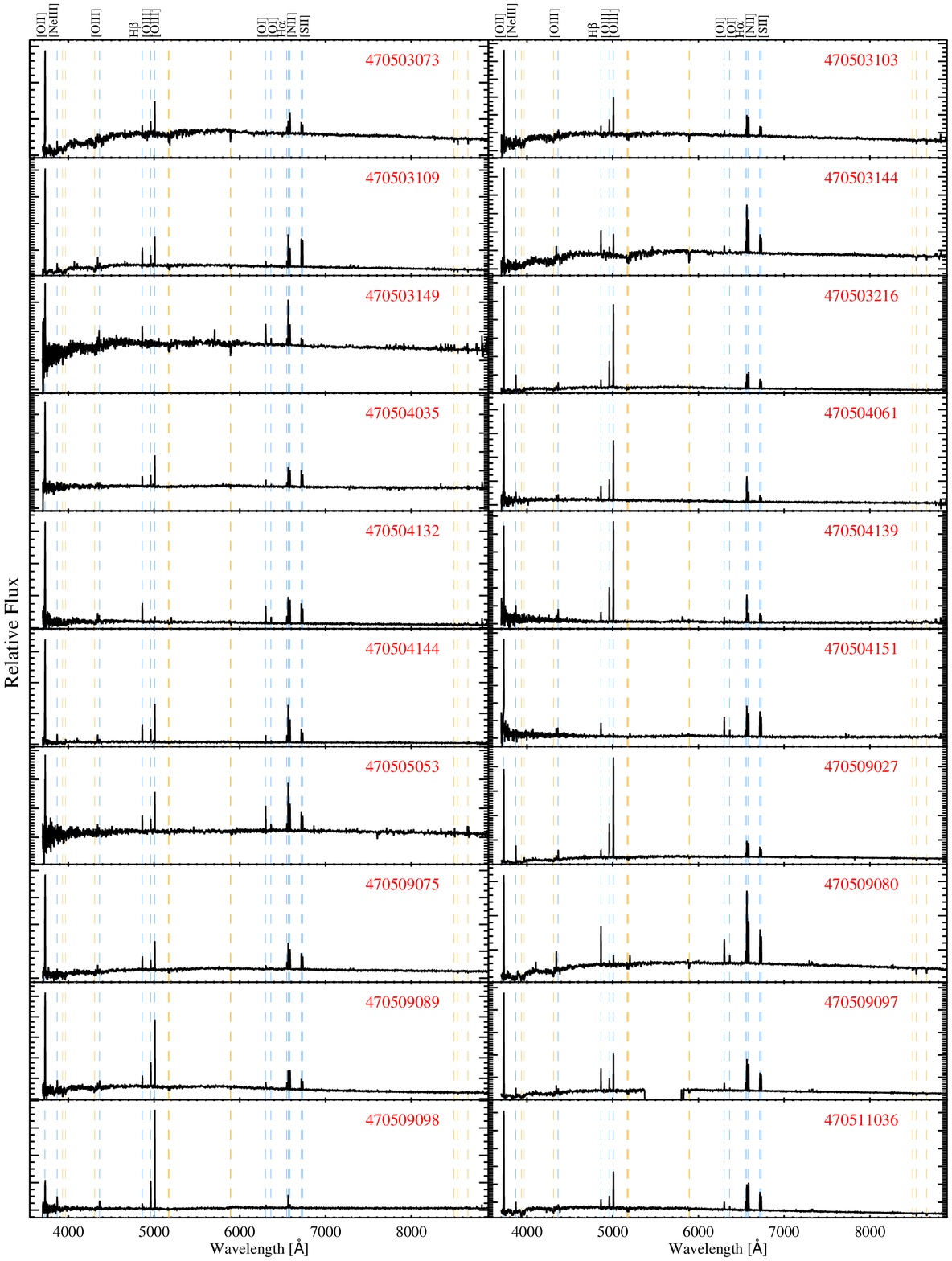}
\figsetgrpnote{Full LAMOST spectra of Group I. Major emission lines are identified with labels (blue lines). Several stellar absorption features are also marked (orange lines). }
\figsetgrpend

\figsetgrpstart
\figsetgrpnum{1.2}
\epsscale{1.1}
\figsetplot{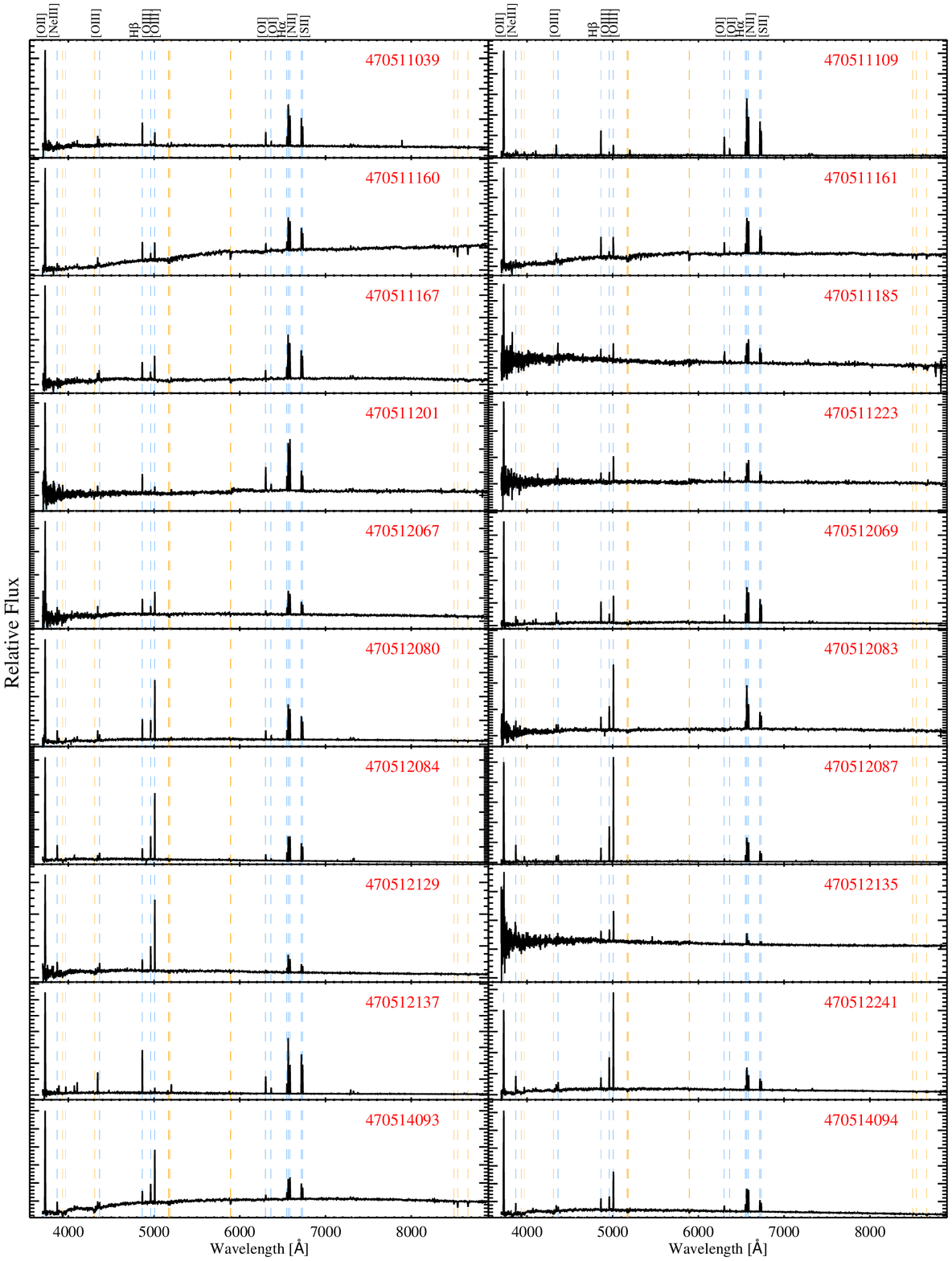}
\figsetgrpnote{Full LAMOST data of Group I spectra with Obs. ID }
\figsetgrpend

\figsetgrpstart
\figsetgrpnum{1.3}
\epsscale{1.1}
\figsetplot{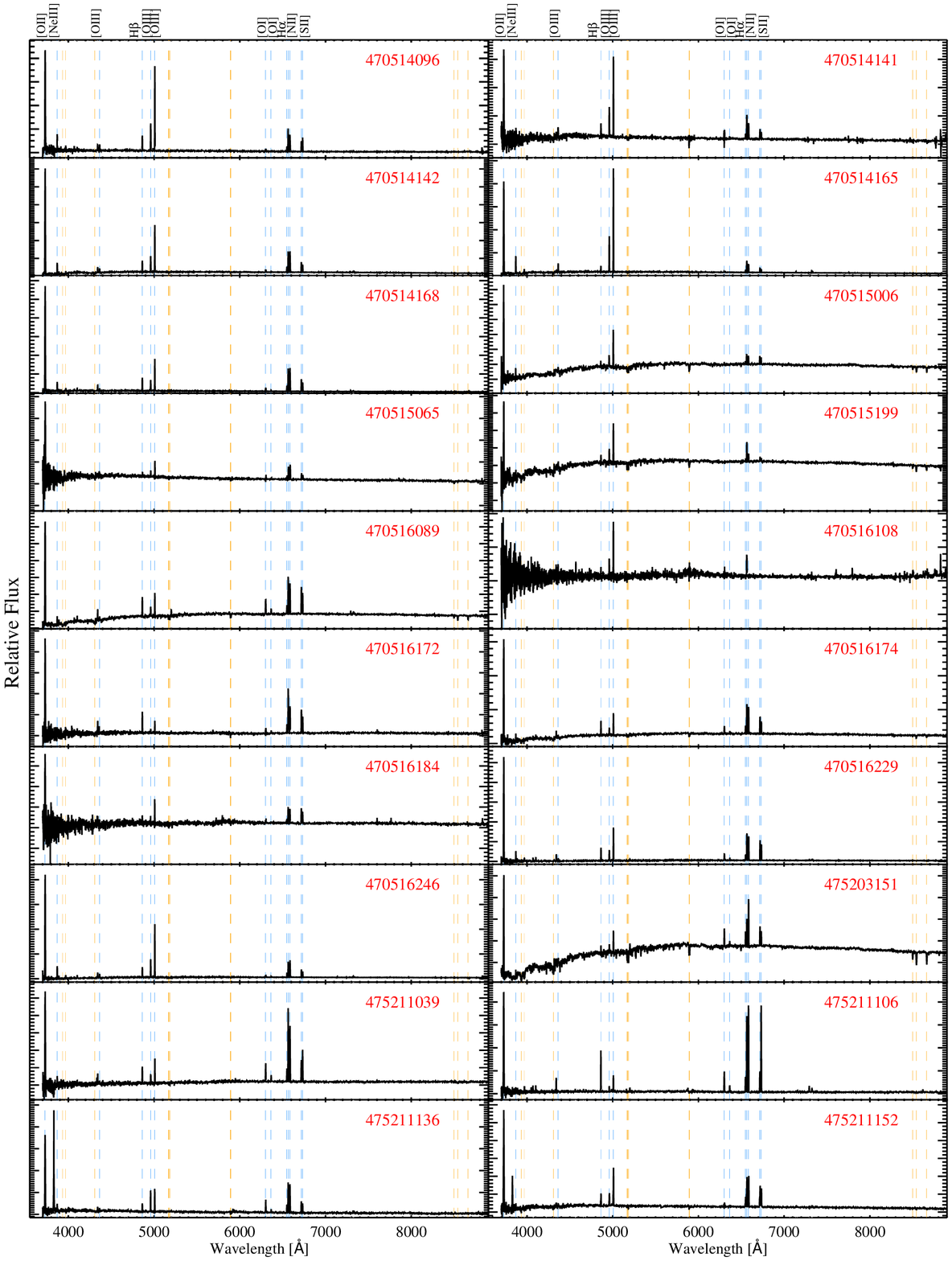}
\figsetgrpnote{Full LAMOST data of Group I spectra with Obs. ID }
\figsetgrpend

\figsetgrpstart
\figsetgrpnum{1.4}
\epsscale{1.1}
\figsetplot{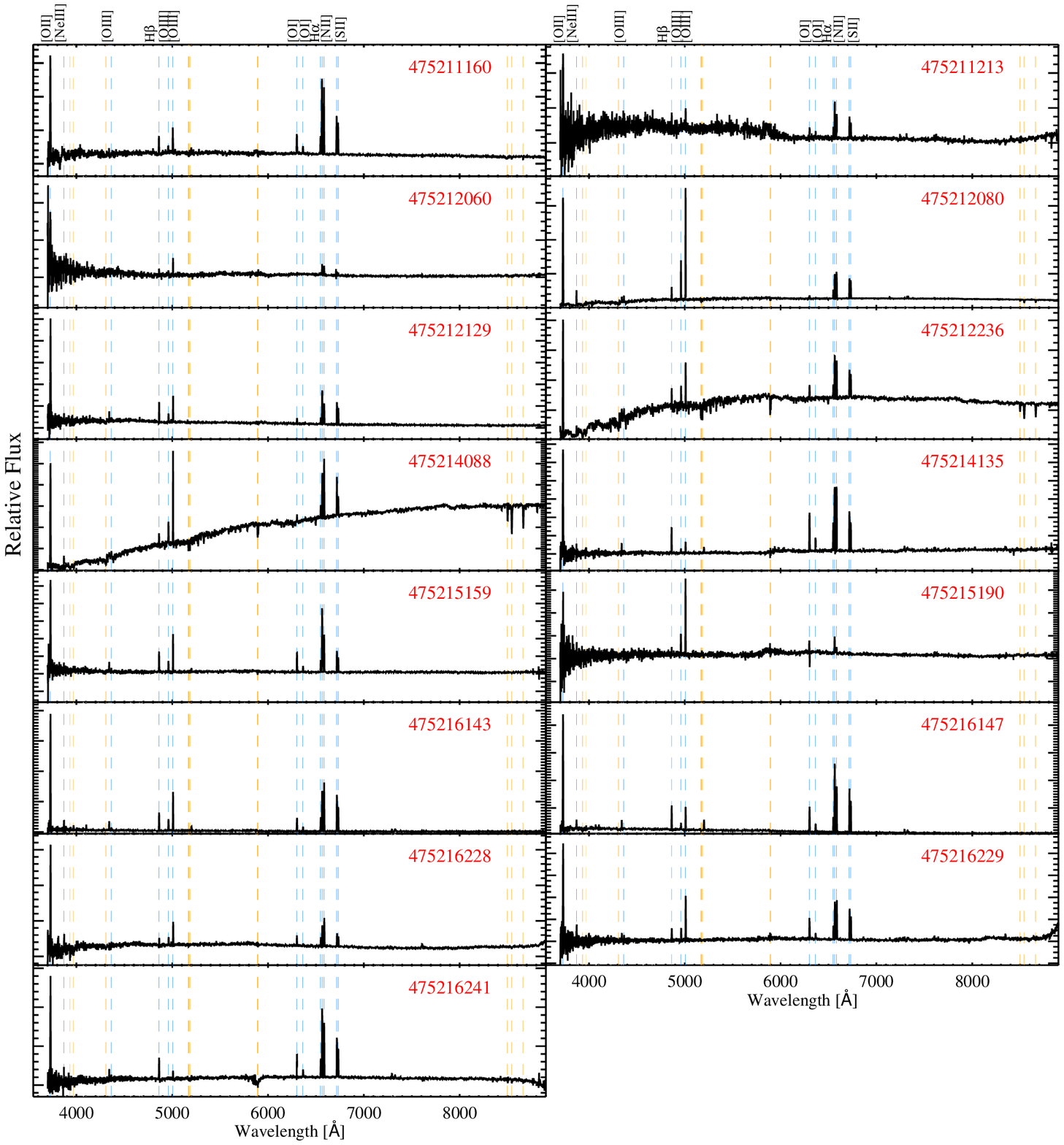}
\figsetgrpnote{Full LAMOST data of Group I spectra with Obs. ID }
\figsetgrpend

\figsetend

\begin{figure}
\figurenum{A. 1}
\plotone{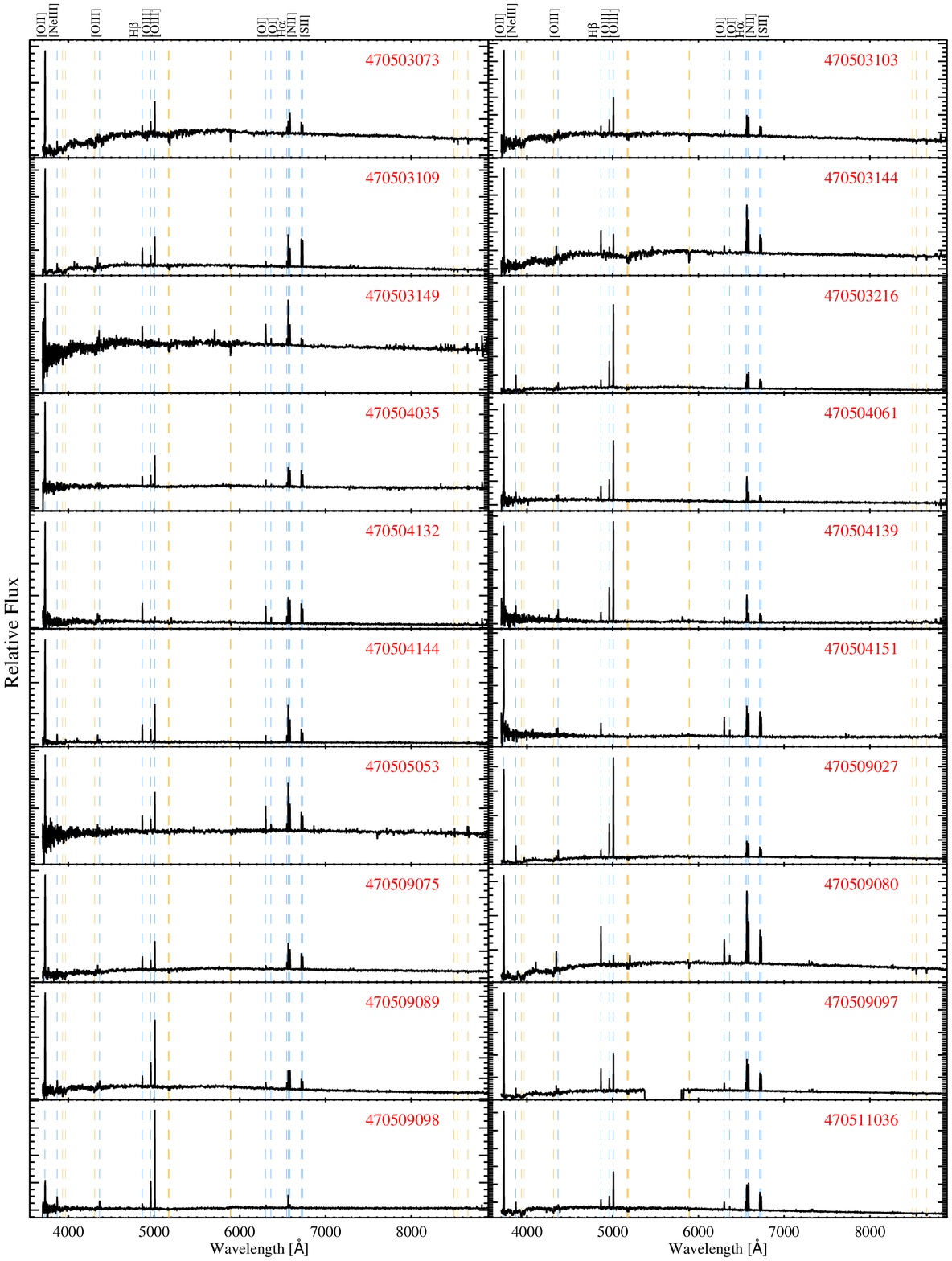}
\caption{ LAMOST data of 75 Group I spectra at the entire wavelengths (3500--8900 \AA). Obs. ID is marked in each panel. Major emission lines are identified with labels at top (blue lines), and several stellar absorption features are also marked (orange lines). The complete figure set (4 images) is available in the online journal.} 
\end{figure}

\end{document}